\documentclass[12pt,preprint]{aastex}
\usepackage{lscape}

\newcommand\hahb{${\rm H}\alpha/{\rm H}\beta$}
\newcommand\nelec{$n_\mathrm{e}$}
\newcommand\ha{${\rm H}\alpha$}

\newcommand\hb{${\rm H}\beta$}

\newcommand\hg{${\rm H}\gamma$}
\newcommand{\grsim}{\mathrel{\hbox{\rlap{\hbox{\lower4pt\hbox{$\sim$}}}\hbox{$>$}}}}

\shorttitle{Fluorescent excitation of Balmer lines}
\shortauthors{Luridiana et al.}

\begin{document}

\title{Fluorescent excitation of Balmer lines in gaseous nebulae: \\ case D}

\author{V. Luridiana}
\affil{Instituto de
Astrof\'{\i}sica de Andaluc\'{\i}a (CSIC), Camino bajo de Hu\'etor 50,
18008 Granada, Spain}
\email{vale@iaa.es}

\author{S. Sim\'on-D\'\i az}
\affil{Observatoire de Gen\`eve and LUTH, Observatoire de Paris, Site de Meudon}
\email{sergio.simon-diaz@obs.unige.ch}

\author{M. Cervi\~no and R.~M. Gonz\'alez Delgado}
\affil{Instituto de
Astrof\'{\i}sica de Andaluc\'{\i}a (CSIC), Camino bajo de Hu\'etor 50,
18008 Granada, Spain}
\email{mcs, rosa@iaa.es}

\author{R. L.  Porter and G. J. Ferland}
\affil{Department of Physics, University of Kentucky, Lexington, KY 40506, USA}

\begin{abstract}
Non-ionizing stellar continua are a potential source of photons for continuum pumping in the hydrogen Lyman transitions. In the environments where these transitions are optically thick, deexcitation occurs through higher series lines. As a result, the emitted flux in the affected lines has a fluorescent contribution in addition to the usual recombination one; in particular, Balmer emissivities are systematically enhanced above case B predictions. The effectiveness of such mechanism in H{\sc~ii} regions and the adequacy of photoionization models as a tool to study it are the two main focuses of this work.

We find that photoionization models of H{\sc~ii} regions illuminated by low-resolution ($\lambda/\delta\lambda\lesssim 1000$) population synthesis models significantly overpredict the fluorescent contribution to the Balmer lines;
the bias has typical values of the order of a few hundredths of a dex, with the exact figure depending on the parameters of the specific model and the simulated aperture.
Conversely, photoionization models in which the non-ionizing part of the continuum is omitted or is not transferred significantly underpredict the fluorescent contribution to the Balmer lines, producing a bias of similar amplitude in the opposite direction.

Realistic estimations of the actual fluorescent fraction of the Balmer intensity require photoionization models in which the relevant portion of the stellar continuum is adequately represented, i.e. its resolution is high in the region of the Lyman lines. In this paper, we carry out such estimation and discuss the variations to be expected as the simulated observational setup and the stellar population's parameters are varied. 
In all the cases explored, we find that fluorescent excitation provides a significant contribution to the total Balmer emissivity. We also show that differential fluorescent enhancement may produce line-of-sight differences in the Balmer decrement, mimicking interstellar extinction.
Fluorescent excitation emerges from our study as a small but important mechanism for the enhancement of Balmer lines. As such, we recommend to take it into account in the abundance analysis of photoionized regions, particularly in the case of high-precision applications such as the determination of primordial helium. 

\end{abstract}

\keywords{atomic processes --- ISM: abundances --- H{\sc~ii} regions --- line: formation --- stars: atmospheres}

\section{Introduction}

The systematic study of gaseous nebulae started in the decade of 1930's  with a long series of papers by Menzel and collaborators \citep[e.g.,][among others]{M37,MB37,BM38}. 
In these papers, the basis was laid for the quantitative analysis of the hydrogen spectrum and the determination of physical conditions in gaseous nebulae.
The hydrogen spectrum was recognized to form in the radiative cascade following recombination of H$^+$ atoms, and 
quantitative predictions for the line intensities based on this scenario were made. Two limiting cases were considered, which depend on the optical thickness of the nebula in the Lyman lines. In case A, the nebula is assumed optically thin in all recombination lines, including those of the Lyman series;
in case B, the nebula is assumed thick to Lyman radiation, which is consequently scattered until conversion into higher series photons plus Ly$\alpha$ or 2-$\gamma$ continuum emission takes place.
Case B was soon shown to be the most appropriate description for most nebulae, both on observational and theoretical grounds \citep[e.g.,][]{BM38,P60}.

A third scenario, named case C, was also initially introduced to represent the case of optically thin nebulae with stellar continuum pumping as an additional mechanism for the excitation of Ly$\alpha$. Contrarily to cases A and B, however, case C has barely had any follow-up in the literature, with very sporadic exceptions \citep[e.g.,][]{C53,F99}.
As pointed out by \citet{F99}, a reason for this scarce popularity is the {\it a priori} assumption of the relative lack of photons at the relevant frequencies due to the absorption features of stellar atmospheres, which would make the mechanism inefficient and imply that case C might be a merely academical scenario with no actual physical applications. (The initial calculation by Baker \& Menzel did not suffer from such limitation, since the stellar source was approximated by a blackbody.)

A further reason for the oblivion of case C is that the scenario it modifies (case A) is inadequate to represent most detectable objects. In other words, the lack of interest toward case C might not have to do with the correction it applies to case A, but rather with case A itself; after all, that nebulae {\sl are} irradiated by stellar light is beyond doubt. If this is the case, one might wonder why not considering the effect of stellar irradiation as a correction to case B, rather than case A.
Since case C is the irradiated counterpart to case A,
it seems natural to christen  ``case D'' the irradiated counterpart to case B  (Table~\ref{tab:cases}).
Note that cases A and B depend solely on the physical conditions in the gas, while cases C and D also depend on the illuminating source, so they are not cases in the same `local' sense of cases A and B.

Case D has never been taken, to the best of our knowledge, into consideration. Although this might seem somewhat surprising at first glance, we believe that there are reasons for it; the arguments developed in this paper will help to understand why, so we defer a discussion of this point to the last section.

\begin{table}
\begin{center}
\caption{Nomenclature for different asymptotic behaviors (``cases'') of the hydrogen spectrum in gaseous nebulae. Column describers  specify the optical depth in the Lyman lines, row describers specify whether irradiation by stellar continuum is included or not.\label{tab:cases}}
\begin{tabular}{l|cc}
\tableline\tableline
 & thin & thick \\
\tableline
Non-irradiated & A & B \\
   Irradiated & C & D \\
\tableline
\end{tabular}
\end{center}
\end{table}

In Section~\ref{sec:mechanism}, we will argue that case D is a better representation of real nebulae than case B, the difference between the two scenarios being mostly small but substantial by modern standards of accuracy. 
Curiously enough, case D is implicitly assumed in many (but not all) photoionization models, although in a biased way; the bias can only be eliminated by feeding into the models high-resolution near-UV spectra (Section~\ref{sec:synthesis}). We will also show that the differential fluorescent contribution to the various Balmer lines can also affect the intrinsic Balmer decrement, pushing it toward either the blue or the red (Section~\ref{sec:spatial}). The resulting spatial distribution of the Balmer decrement depends on the spatial distribution of stars and the gas around them. A brief discussion of the dependence of fluorescence enhancement on the nebular and the stellar population's parameters is given in Section~\ref{sec:dependencies}, and on details of the assumed stellar atmospheres 
is given in Section~\ref{sec:atmospheres}.
The main focus of Section~\ref{sec:discussion} is the comparison of fluorescence enhancement to the other major deviation to case B, namely collisional excitation of H$^0$. Finally, a short summary of the paper is given in Section~\ref{sec:summary}.

\section{The mechanism}\label{sec:mechanism}

The abundance analysis of gaseous nebulae is based on a very simple principle.
The intensity of a line with respect to, e.g., \hb\ is given by the abundance of the emitting ion relative to H$^+$ times a mathematical expression describing the physical mechanisms of line formation involved. If such mechanisms are known and all the parameters they depend on (such as the electronic temperature $T_e$) can be determined, then the observed value of a line ratio can be used to measure the ionic abundance relative to hydrogen. 

The specific mechanisms involved depend on the ion and the regime considered. To mention but a few examples, it is known that He I lines in H{\sc~ii} regions mainly form in the recombination cascade of He$^+$, plus a density-dependent contribution by collisional excitation from the metastable level 2$^3$S, which affects mainly the triplet lines \citep{OF06}. He{\sc~ii} lines form in the recombination cascade of He$^{++}$. Heavy element lines mainly form through collisional excitation from the ground state of abundant ionic stages. Finally, H{\sc~i} lines - a standard reference for the intensity of the other lines in the optical range - form mostly in the radiative cascade following recombination of H$^+$ under case B conditions.

It is also known that case B is only a limiting case and not the exact description of conditions in H{\sc~ii} regions; strictly speaking, it is not even self-consistent, as it provides no mechanism to depopulate the 2{\it p} level \citep[see, e.g.,][]{B58}.
However, the predictions made by assuming case B have generally proved to be accurate, and only in recent times has it become possible (and relevant) to point out exceptions to it. One such exception is the occurrence of collisional excitation from the ground state, which has recently been rediscovered and closely studied as a consequence of the increasing need for accuracy in abundance determinations \citep[][]{SI01,Lal03,PLP07,L50}. 
A further way in which hydrogen may deviate from case B behavior, which is the subject of this paper, is continuum pumping in the Lyman lines and the related Balmer fluorescent enhancement.

Continuum pumping is the excitation of a transition by means of the absorption of a stellar continuum photon of the appropriate wavelength. Such excitation is followed by a radiative cascade which, in optically thick environments such as those of H{\sc~ii} regions, ends up producing a Balmer series photon plus either Ly$\alpha$ or 2-$\gamma$ continuum. In such a way, a Ly$\beta$ absorption at 1026 \AA\ can lead to the emission of an H$\alpha$ photon; a Ly$\gamma$ absorption at 973 \AA\ can lead to the emission of either H$\beta$ or Pa$\alpha$ plus H$\alpha$; and so on. These Balmer photons appear independently of, and in addition to, Balmer photons formed in the recombination cascade of H$^+$, so these two contributions add up in the total spectrum. As for Ly$\alpha$ photons, they are scattered and gradually shifted in frequency until they escape the nebula, or are absorbed by dust; in dustless nebulae, a significant fraction of Ly$\alpha$ photons is converted into 2-$\gamma$ continuum. Dust can also destroy higher Lyman photons, thus inhibiting pumping; this possibility will be briefly discussed in Section~\ref{sec:discussion}.

It must be stressed that this Balmer enhancement mechanism is triggered by pumping at the Lyman-line wavelengths, i.e. it depends on the shape of the {\it non-ionizing} stellar field. This runs counter the common wisdom that, in a photoionized object, only the ionizing part of the stellar field matters; after all, the ionization structure is governed by the ionizing radiation, and the temperature structure is determined by the balance between heating due to photoionization and cooling due to collisional excitation of abundant metals and, in less metal-rich environments, Ly$\alpha$ emission. As a consequence, in photoionization modeling the effect of non-ionizing stellar radiation on the nebula is generally neglected. 
Yet, the emission spectrum of an H{\sc~ii} region can be altered by continuum pumping even if this mechanism has no other effect on the structure of the nebula. In the next section, we will justify this statement by discussing the role of stellar continuum pumping in case C nebulae, referring in particular to \citeauthor{F99}'s \citeyearpar{F99} results. The following sections will be dedicated to the role of stellar pumping in the optically thick case, which we have labelled above ``case D''. In particular, we will estimate the contribution of stellar pumping\footnote{A note on nomenclature: although continuum pumping and fluorescent excitation are different processes (the former referring to level excitation, the latter to line emission), they are linked together, so we will use the two terms interchangeably whenever appropriate.} on Balmer intensities and show that it cannot be neglected.

\subsection{Case C}\label{sec:caseC}

A good starting point to case D is case C, which applies to nebulae optically thin in the Lyman lines irradiated by a stellar continuum. In an optically thin environment, stellar continuum pumping results in a significant enhancement of Ly$\alpha$ emission, sufficient to bring the Ly$\alpha$/\hb\ ratio well above the case B value, rather than below it as predicted for case A. \citet{F99} explains this by showing that, in the vicinity of the star (i.e., before stellar Lyman-line photons have substantially been absorbed) continuum pumping provides a source of Balmer photons comparable to recombination. If the gas slab is optically thin, recombination and fluorescence act concurrently in the whole volume, so the total emissivity in Ly$\alpha$ is largely enhanced by the latter mechanism.

This argument anticipates one reason why case D has never been taken into consideration: if the gas thickness is large (i.e. the optical depth is large in the Lyman continuum), Lyman-line photons are used up much earlier than Lyman continuum photons, so that the relative effect of continuum pumping on the total emitted spectrum is diluted; the larger the optical depth, the more diluted the effect, until it possibly becomes negligible altogether. If this were always the case, case D would not have any actual applications.
There are, however, three possible exceptions to this conclusion:
a) a high accuracy in abundance analysis is required; 
b) the nebular region affected by fluorescence has an unusually large weight in the observation, due to aperture effects;
c) the balance between the Lyman-line photons and the Lyman-continuum photons is more in favour of the former than in the case considered by \citeauthor{F99}.

\begin{figure}
\plotone{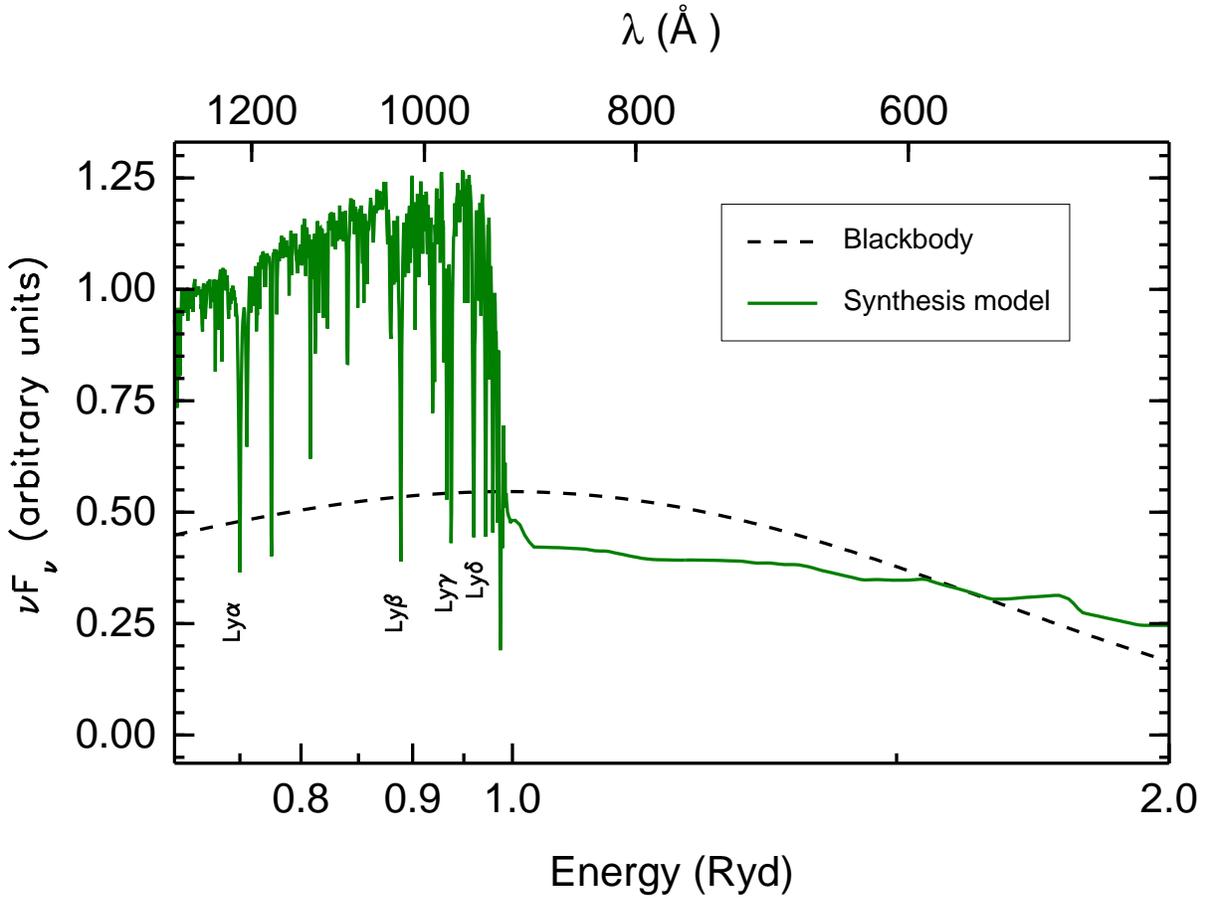}
\caption{Comparison between the SEDs of a population synthesis model and a blackbody at $T_\mathrm{eff} = 40$ kK, normalized to the same number of ionizing photons. The position of the first four Lyman lines is also shown.\label{fig:synthesis_vs_blackbody}}
\end{figure}

The first circumstance is certainly met in primordial helium abundance determinations, a goal that requires knowing the line formation mechanism of H and He lines with an accuracy better than 1\%.
The second circumstance is met in spatially-resolved observations sampling small regions in the vicinity of the ionizing cluster; observations of this kind are more and more common as technology improves.
The third circumstance does not seem at first glance very probable when considering that the input spectrum of \citet{F99} is a blackbody; these results, one may be tempted to conclude, overestimate rather than underestimate the amount of continuum pumping, since a blackbody has no absorption features. Although this is true, it must be considered that a blackbody has no Lyman jump either, so the Lyman line-to-continuum flux of \citet{F99} might actually be an underestimate. To demonstrate this point, in Figure~\ref{fig:synthesis_vs_blackbody} we compare the $T_\mathrm{eff}=40,000$ K blackbody of that paper to the synthetic spectrum of a $t=3$ Myr population emitting the same rate of ionizing photons (i.e., with the same potential to produce \ha\ photons 
by recombination; more details on this spectrum will be given in Section~\ref{sec:caseD}). The figure shows that the blackbody does not lie {\sl above} the synthesis spectrum: rather, it crosses the near-bottom of the relevant absorption lines, thus underestimating the number of photons capable of continuum pumping.

Concluding, we see that there is no {\it a priori} reason to discard the relevance of case D. In the following section, we will quantify in a specific case the correction represented by case D with respect to case B, and discuss how this result is expected to depend on the relevant H{\sc~ii} region's parameters.

\subsection{Case D}\label{sec:caseD}

\begin{figure}
\plotone{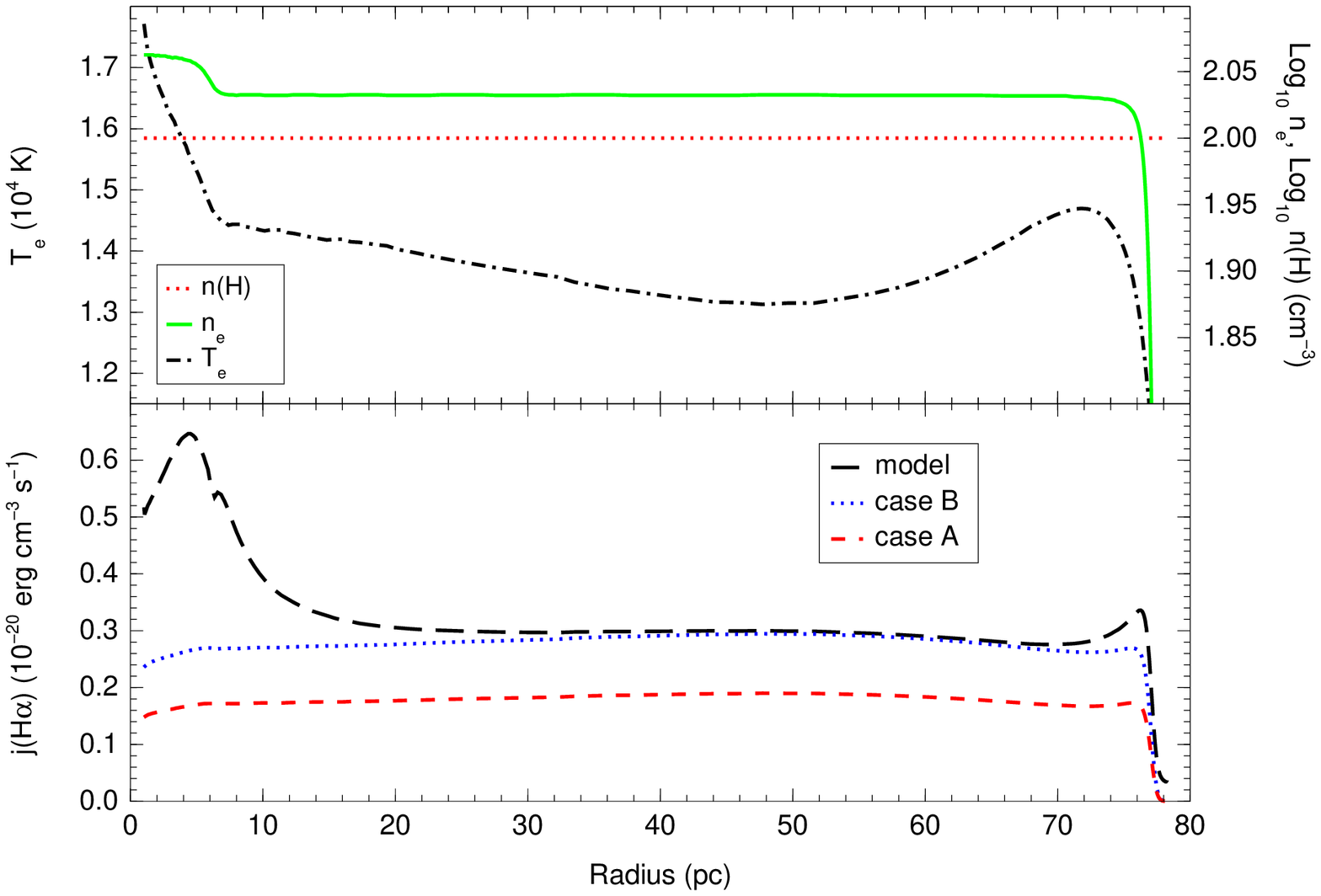}
\caption{Upper panel: physical conditions in a model low-metallicity giant H{\sc~ii} region ionized by a 3 Myr stellar population (see Section~\ref{sec:mechanism} for details). Lower panel: normalized \ha\ emissivity profile of the model, compared to case A and B predictions.\label{fig:NeHdenTeHb_M1}}
\end{figure}

The upper panel of Figure~\ref{fig:NeHdenTeHb_M1} shows the physical conditions
in a model giant H{\sc~ii} region, named M1, computed with Cloudy \citep[version C08.00;][]{Pal08}. This version of the code includes a full model of the hydrogen atom with strong radiation field physics \citep[][and Appendix~\ref{sec:appendix}  of this paper]{FR88,FF97,Pal08}. For the sake of simplicity, we assumed a dustless nebula with a simple spherical geometry, with starting radius $R_{in}=1$ pc, hydrogen number density $N_H=100$ cm$^{-3}$ and filling factor $\epsilon=0.01$. Although a blister model would be more appropriate to represent many, if not most, H{\sc~ii} regions, we prefer to stick to the simplest possible case. The volume-integrated predictions presented in this paper do not depend on the assumed geometry, while the spatially-resolved predictions do; it is therefore impossible to give figures suitable for any geometry and any line of sight, so the reader is warned that our spatially-resolved predictions are given more as a rough indication of the magnitude of the effect than as a definitive estimate. The model's total metallicity is $Z_{gas}=0.001$, and the detailed elementary abundances are given in Table~\ref{tab:chem}; dust is not important at such low metallicities. Thermal line widths are assumed; the influence of a non-thermal contribution is discussed in Section~\ref{sec:turbulence}. The ionizing source is the spectral energy distribution (SED) of a $t=3$ Myr stellar population, emitting ionizing photons at a rate Log $Q(\mathrm{H}^0)$ = 52.09 s$^{-1}$. the SED has been computed with Sed@ \citep[][and references therein]{CL06}, and does not include stellar winds. The effect of stellar winds on the predictions is discussed in Section~\ref{sec:atmospheres}.
 
Assuming that the model's emission is described by case B, a fairly constant \ha\ emissivity profile is expected, given the shallow temperature gradient. This, however, is not the case: two exceptions to case B behavior can be seen in the \ha\ profile (Figure~\ref{fig:NeHdenTeHb_M1}, bottom panel).

\begin{table}
\begin{center}
\caption{Elementary chemical abundances in the reference model, given as 12 + Log(X/H). The remaining elements have not been included in the calculation since they are quite unimportant.\label{tab:chem}}
\begin{tabular}{lr@{.}l}
\tableline\tableline
& & \\
Element & \multicolumn{2}{c}{Abundance} \\
& & \\
\tableline
He & 10&89 \\
C  &  7&62 \\
N  &  6&38 \\
O  &  7&87 \\
Ne &  7&15 \\
S  &  6&47 \\
Cl &  4&48 \\
Ar &  5&50 \\
Fe &  5&81 \\
\tableline
\end{tabular}
\end{center}
\end{table}

In the external part ($r \grsim$ 70 pc) there is a small bump in the emissivity profile: this bump is caused by collisional excitation of neutral hydrogen, a mechanism ignored in case B and which only becomes effective in the outer part, thanks to the drop in the hydrogen ionization fraction. A very low metallcity gas,
which we consider here, will have a high kinetic temperature so these
collisional processes are especially important; more details on this can be found in \citet{L50}.

In the inner part there is a much larger bump, extending through a substantial fraction of the total radius and whose maximum value more than doubles the expected case B value, which is shown as a dotted line. When the upper and lower panels of Figure~\ref{fig:NeHdenTeHb_M1} are simultaneously considered, it becomes clear that density and temperature cannot be responsible for the excess emissivity, as one can find pairs of points with the same $T_e$ and \nelec\ values but very different emissivities. In other words, the cause of this behavior cannot be sought in the gas: this leaves the radiation field as the only alternative. Indeed, the excess emissivity is caused by fluorescent excitement of \ha\ by continuum pumping of Lyman photons (mainly Ly$\beta$). This is consistent with the position of the bump, close to the ionizing source
(note that this bump has nothing to do with the breaks in the \nelec\ and $T_\mathrm{eff}$ gradients seen at 6 pc, which are caused by the recombination of He$^{++}$ and O$^{3+}$).

In the following sections, we will discuss whether the behavior of this model is representative of real nebulae and which its consequences are for the analysis of H{\sc~ii} regions, considering several specific issues.
We will start by considering a simple model of a spherical H{\sc~ii} region, ionized by a population synthesis model (Section~\ref{sec:synthesis}), and show the effect of continuum pumping on the Balmer intensities and ratios.
In the following sections we will complicate this basic scenario by discussing the effect of dust, turbulence, geometry, or a different metallicity (Section~\ref{sec:dependencies}).
Since the amount of continuum pumping depends on the detailed shape of the spectrum, we will also discuss the effect of assuming stellar atmospheres with shocked stellar winds. Since no high-resolution synthesis models are available with shocked winds, we will consider the case of a model H{\sc~ii} region ionized by individual stars (Section~\ref{sec:atmospheres}). Sections~\ref{sec:discussion} and \ref{sec:summary} will be devoted to the discussion and conclusions, respectively.

\section{Hydrogen fluorescent excitation predicted with population synthesis models }\label{sec:synthesis}

\begin{figure}
\plotone{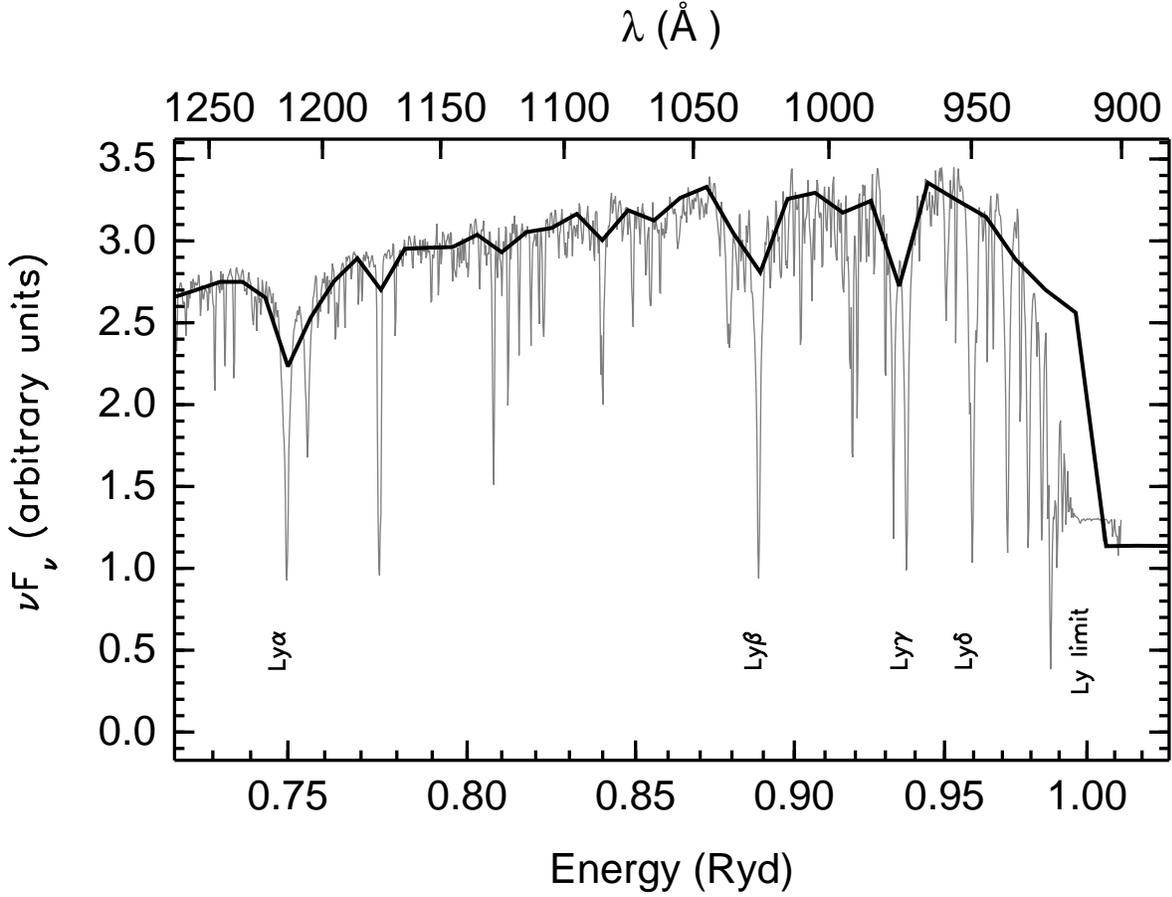}
\caption{UV spectral energy distribution of a 3 Myr, $Z=0.001$ stellar population, in low resolution (thick black line, spectral sampling = 10 \AA) and high resolution (thin grey line, spectral sampling = 0.3 \AA). Also shown are the first four Lyman lines, from Ly$\alpha$ at $\lambda =$ 1215 \AA \ (which does not produce any Balmer photon) through Ly$\delta$ at 950 \AA. \label{fig:SED_UV}}
\end{figure}

The black thick curve in Figure~\ref{fig:SED_UV} shows the near-UV portion of the SED fed into the model described in Section~\ref{sec:caseD}; we stress again that this SED, as all the synthesis models used in this work, has been computed with stellar spectra with no emission lines. In this wavelength range, the spectral sampling
of this SED is $\Delta\lambda=10$ \AA; this value is determined by the grid of stellar libraries used, i.e. it does not depend on the way in which population synthesis is carried out; any synthesis model using the same stellar library would have the same resolution \citep[this is the case, for example, of Starburst99 models:][]{Lal99}.
At such resolution, stellar absorption lines are smoothed out, so that the number of Lyman photons in the spectrum is overestimated. Since the mechanism studied in this work depends directly on the number of Lyman-line photons emitted, it is important to establish whether fluorescent enhancement of \ha\ can actually occur in H{\sc~ii} regions or it is just an artifact of low-resolution synthesis models. The only way to answer this question is by feeding a high-resolution SED into the photoionization code.

\subsection{Photoionization models with high resolution SEDs}\label{sec:photo_models_HR}

To address the issue mentioned in the previous section, we computed a
new photoionization model, labelled M2, identical to M1 in all
respects except for the stellar source, which has been replaced with a
high-resolution SED in the 900-2000 \AA\ range. This SED represents a
population of the same stellar parameters (age and metallicity) as
those of the original SED, but has a much better spectral sampling
$\Delta \lambda$ = 0.3 \AA \
in the near-UV range; a portion of this spectrum is shown in
Figure~\ref{fig:SED_UV} (thin grey curve).
The SED is taken from a library of high-resolution 
synthesis models computed with the code Sed@ \citep{CL06} to carry out this work.
The library has been computed with the non-rotating tracks of the Geneva group
\citep[][]{Sal92,Sal93a,Sal93b,Cal93},
including standard mass-loss rates computed at five metallicities:
$Z$=0.040, 0.020, 0.008, 0.004 and 0.001. The isochrones are
calculated by means of a parabolic interpolation between the tracks
(log $M$, log $t_k$) in the HR diagram on a variable mass grid as
prescribed in \cite{Cal01}. The isochrones were generated at
specific ages with a time step $\Delta t=1$ Myr for 1-10 Myr. We have
assumed a power-law IMF, with slope ${\alpha}=-2.35$ \citep{S55} and
low- and high-mass cutoffs $M_{\rm low}=0.1\, M_\odot$ and $M_{\rm
up}=120\, M_\odot$, respectively.

To model the SED, we have combined two different sets of atmosphere models, 
with low and high resolution respectively.
Low resolution atmosphere models are standard
Kurucz LTE atmospheres.
High resolution atmosphere models cover the
wavelength range from 900 to 2000 \AA, and have been computed using the same
atmospheres as the ones described in \cite{Mal05a,GDal05} but in
the UV domain.  Hot stars ($27500\leq T_{\rm eff}\leq 55000$ K) are 
modelled as line-blanketed, non-LTE, plane-parallel, hydrostatic atmospheres 
\citep{LH03}, and stars with $8000\leq T_{\rm eff}\leq27000$ K are
modelled with Kurucz LTE atmospheres \citep[][]{K93}. 
We have assumed no
contribution in the UV from stars with lower temperatures. The 
high-resolution synthetic spectra are obtained with the program SYNTHE by
\cite{HLJ95}. 

The portions of the SED shortward of 900 \AA\ and longward of 2000 \AA\ are equal in the low- and high-resolution models. 
The main features of these models are given in Table~\ref{tab:models}, together with those of the models presented in the remainder of the paper.

A summary of the difference between M1 and M2 from the point of view of the fluorescent contribution to \ha\ is given in Table~\ref{tab:HRvsLR}. The three rows in the table correspond to different ways of integrating the emissivity along the shell structure of the model: in the 1D integration the emissivity of each shell is weighted by the thickness of the shell $dr$; in the 2D integration it is weighted by the ring surface $2\pi r\,dr$; and in the 3D integration it is weighted by the shell volume $4\pi r^2\, dr$. Observationally, the three cases correspond to a beam-like aperture, a narrow slit, and an aperture including the whole object respectively.
Since fluorescent excitation takes place in the inner part of the nebula, we expect that the effect on integrated luminosities of differences in the local emissivity is amplified in 1D integrations with respect to 3D integrations.
As expected, Table~\ref{tab:HRvsLR} shows that \ha\ is spuriously enhanced in M1 with respect to M2 as a consequence of the poor resolution of its input spectrum, which partially suppresses stellar absorption lines. The bias amounts to as much as 0.04 dex (9\%) in the case of a beam-like simulated aperture, relative to the {\sl total} H$\alpha$ intensity. 

\begin{landscape}
\begin{table}
\begin{center}
\caption{Main features of the photoionization models used in this work.\label{tab:models}}
\begin{tabular}{cccccccc}
\tableline\tableline
Model & \multicolumn{3}{c}{Wavelength range (\AA)} & {Nebular boundary} & \multicolumn{3}{c}{Excitation mechanism$^a$} \\
& $>$ 2000 & 2000 - 900 & $<$ 900 & & F & R & CE \\
\tableline
M1 & LR & LR & LR & Radiation bounded & + & = & = \\
M2 & LR & HR$^b$ & LR & Radiation bounded & = & = & = \\
M3 & LR & suppressed$^c$ & LR & Radiation bounded & $-$ & = & = \\
M4 & LR & HR$^b$ & LR & Density bounded & = & = & $-$ \\
M5 & LR & suppressed$^c$ & LR & Density bounded & $-$ & = & $-$ \\
\tableline
\end{tabular}
\tablenotetext{a}{F=fluorescence; R=recombination; CE=collisional excitation; the symbols `+', `=' and `$-$' denote that the mechanism is overestimated, properly included, and underestimated respectively.}
\tablenotetext{b}{Sampled to $\Delta\lambda/\lambda=0.0005$  (see Section~\ref{sec:photo_models_HR}).}
\tablenotetext{c}{In the 910-1030 \AA \ range.}
\end{center}
\end{table}
\end{landscape}

\begin{table}
\begin{center}
\caption{Bias in the total \ha\ intensity predicted with a low-resolution stellar continuum. The meaning of labels 1D, 2D and 3D is explained in Section~\ref{sec:photo_models_HR}.\label{tab:HRvsLR}}
\begin{tabular}{ccc}
\tableline\tableline
& & \\
Integration & $\frac{I(\mathrm{H}\alpha)_\mathrm{M1}}{I(\mathrm{H}\alpha)_\mathrm{M2}}$ -1 \\
& & \\
\tableline
1D & 0.09 \\
2D & 0.03 \\
3D & 0.02 \\
\tableline
\end{tabular}
\end{center}
\end{table}

Since predicting Balmer line intensities is not a common application of photoionization models, which is the reason for pointing out this bias?
The main reason to do it is estimating the amount of fluorescent excitation in real objects so that it can be taken into account when chemical abundances are determined. The exercise realized in this section provides a necessary piece of information in order to know whether such estimation can be carried out with low-resolution models, or high-resolution models are needed instead. The results above indicate that the bias in \ha\ introduced by low-resolution models is substantial if compared to our best possible guess of the total emitted \ha\ intensity, which is given by M2. Therefore, the results of Table~\ref{tab:HRvsLR} indicate that low-resolution synthesis spectra are not suitable to study the effect of fluorescence, and that high-resolution spectra are needed instead; they also indicate that the predicted spatially-resolved line ratios relative to \hb\ predicted are grossly overestimated in the direction of the center; this must be taken into account when photoionization models are used to fit spatially-resolved spectroscopic data, as done, e.g., in \citet{LP01}.

\begin{figure}
\epsscale{.50}
\plotone{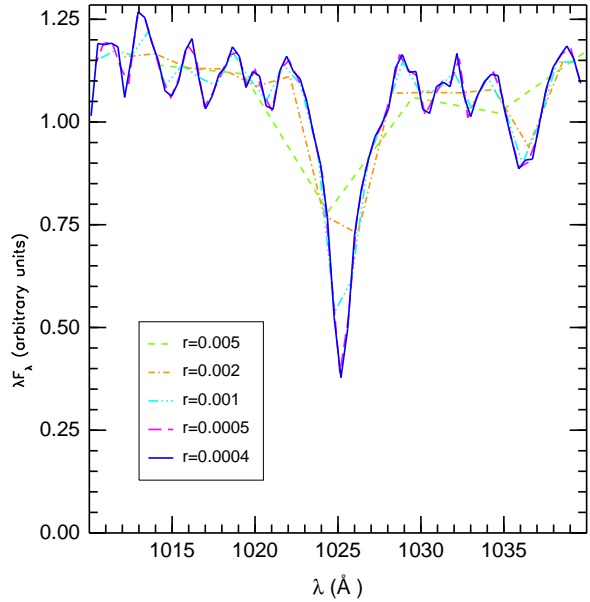}
\caption{HR synthesis model rebinned by Cloudy to different values of the spectral resolution $r=\Delta \lambda/\lambda$.\label{fig:comparing_rCloudy_UV}}
\end{figure}

Note that Cloudy automatically rebins any input spectrum, so that to take advantage of the larger spectral resolution it has been necessary to increase Cloudy's default continuum mesh, as explained in the code's documentation (Hazy, part 3, page 410). We carried out several different experiments in this direction exploring the resolution range $0.005 \le \Delta \lambda/\lambda \le 0.0004$, corresponding to spectral samplings $5 \mathrm{\AA} \ge \Delta \lambda \ge 0.4 \mathrm{\AA}$ at $\lambda =$ 1000 \AA. Some of the spectra used internally by Cloudy after such rebinning are shown in Figure~\ref{fig:comparing_rCloudy_UV} in the Ly$\beta$ wavelength region. We found no significant differences in the emission-line spectrum below $\Delta \lambda/\lambda \lesssim 0.0005$. 

As a final comment, we note that, since real massive stars may develop P-Cygni line profiles due to the presence of strong stellar winds, at very high spectral resolution one would also need to consider both the detailed line profile and the relative velocities of the star and the nebula, which determine which part of the spectrum fall at the relevant wavelengths.

\subsection{Net contribution of fluorescent excitation to Balmer lines}\label{sec:fluo_contribution}

Having established that low-resolution spectra are unsuitable to compute the fluorescent contribution to Balmer lines, we go on to address this problem 
by means of high-resolution spectra, considering the specific case of model M2.
The fluorescent intensities in M2 can be estimated by computing a new model M3, identical to M2 in all respects except for fluorescent excitation, which is suppressed by setting to zero the relevant portion of the continuum (but it can also be switched off by the Cloudy command ``no induced processes''). 
The fluorescent enhancement of M2 is computed as the difference between the Balmer line intensities of M2 and M3 (Table~\ref{tab:HR2vsNoPumping}). For the reasons explained in Section~\ref{sec:photo_models_HR}, the enhancement is particularly large in the beam-like integration: specifically, in our models {\it as much as  10\% of the total intensity in H$\alpha$ and 14\% in H$\beta$  seen in the direction of the nebular center is caused by fluorescence rather than recombination}.
For completeness, we have also computed the fluorescent enhancement of Balmer lines of model M1, to quantify the fraction of emissivity that would - erroneously - be attributed to fluorescence enhancement, had we used model M1 to this aim: in that case, the estimated fluorescent enhancement is almost twice as large as in the unbiased case (Table~\ref{tab:HR2vsNoPumping}). 
As anticipated in Section~\ref{sec:photo_models_HR}, this demonstrates that inputting low-resolution spectra to photoionization models grossly biases the estimated fluorescence contribution.

Thus, the first general result is that fluorescence does indeed has a sizeable influence on the emitted Balmer spectrum. 
In the remainder of this section we will discuss the implications of such influence. In the following sections we will briefly discuss how our results depend on the model's parameters, such as the geometrical assumptions on the nebular distribution and the parameters that define the stellar populations.

\begin{table}
\begin{center}
\caption{Fraction of total intensity contributed by fluorescent excitation in models M2 and M1.\label{tab:HR2vsNoPumping}}
\begin{tabular}{cccccccc}
\tableline\tableline
Integration & \multicolumn{3}{c}{1-$\frac{I(\lambda)_\mathrm{M3}}{I(\lambda)_\mathrm{M2}}^a$} && \multicolumn{3}{c}{1-$\frac{I(\lambda)_\mathrm{M3}}{I(\lambda)_\mathrm{M1}}^b$} \\
\cline{2-4} \cline{6-8}
& H$\alpha$ & H$\beta$ & H$\gamma$ & & H$\alpha$ & H$\beta$ & H$\gamma$ \\
\tableline
1D & 0.10 & 0.14 & 0.14 && 0.17 & 0.21 & 0.26 \\
2D & 0.03 & 0.04 & 0.05 && 0.06 & 0.06 & 0.09 \\
3D & 0.02 & 0.02 & 0.02 && 0.04 & 0.03 & 0.04 \\
\tableline
\end{tabular}
\tablenotetext{a}{True fluorescent enhancement.}
\tablenotetext{b}{Biased fluorescent enhancement.}
\tablecomments{Total intensities include fluorescence, recombination, and collisional excitation.}
\end{center}
\end{table}

\subsubsection{Effect of fluorescent enhancement on abundance determinations}

The results listed in Table~\ref{tab:HR2vsNoPumping} have important implications for abundance determinations. When a beam aperture is applied to the center of model M3, at least 10\% of the total Balmer intensity comes from fluorescent excitation. If model M3 were a real object analyzed under the assumption of case B emissivity, and assuming that no fluorescent excitation is taking place in other ions, the inferred abundances of helium and heavy elements would be underestimated by up to 0.06 dex. Even in the most favourable case, the large aperture corresponding to a 3D integration, a 0.01 dex (2\%) bias would be introduced; although this is negligible for most abundance measurements, it would be crucially large for high-precision applications such as the determination of primordial helium. Although helium lines might also be affected by fluorescent excitation, line ratios will in general be different from what predicted when neglecting fluorescent excitation.

\subsubsection{Effect of fluorescent enhancement on the Balmer decrement}\label{sec:effect_Balmer_dec}

\begin{figure}
\epsscale{.80}
\plotone{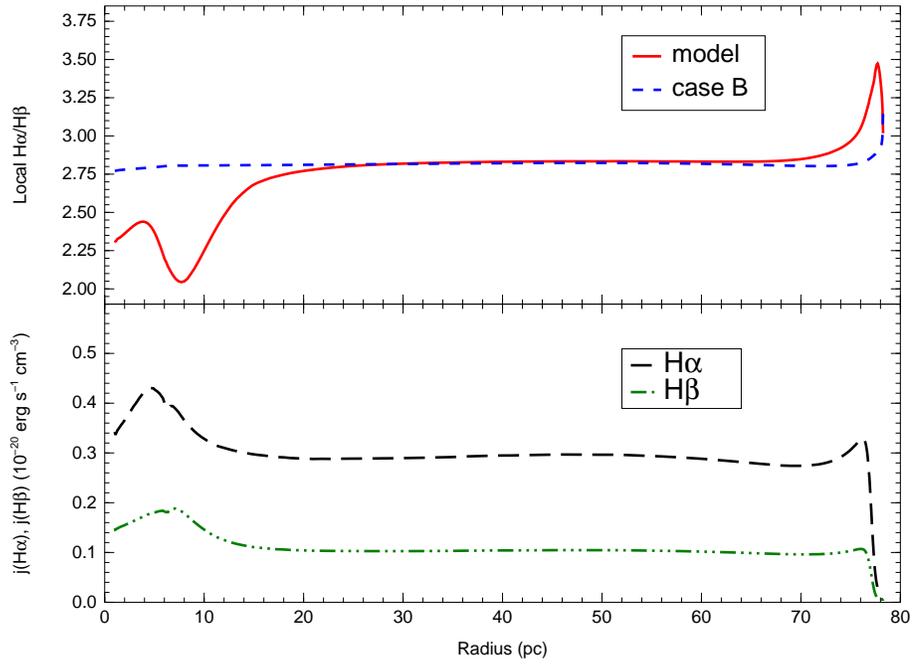}
\caption{Top panel: Local \hahb\ intensity ratio as a function of radius in model M2, compared to the case B value at the model's local temperature. Bottom panel: normalized \ha\ and \hb\ emissivity profiles.\label{fig:Ha2Hb_HaHb_r}}
\end{figure}

A second implication concerns extinction measurements. 
The upper panel of Figure~\ref{fig:Ha2Hb_HaHb_r} shows the local \ha/\hb\ radial profile of model M2. In the innermost 20 pc and the outermost 20 pc, this ratio differs considerably from the value expected under case B. The deviation in the external region is caused by collisional excitation: \hahb\ rises sharply because collisional excitation is more effective for \ha\ than for \hb\ \citep{L50}. The deviation in the internal region is caused by fluorescence, as can be shown by an analysis of the pattern it follows. Fluorescent excitation of \ha\ and \hb\ is mainly triggered by Ly$\beta$ and Ly$\gamma$ photons, respectively; the optical depth in Ly$\beta$ is larger than that in Ly$\gamma$, so that \ha\ peaks at a smaller radius than \hb\ (Figure~\ref{fig:Ha2Hb_HaHb_r}, bottom panel). As a consequence, the \ha/\hb\ ratio initially increases, then it decreases as Ly$\beta$ photons become exhausted and Ly$\gamma$ photons are still available, and eventually increases again to approach case B as both Ly$\beta$ and Ly$\gamma$ photons are exhausted. This behavior is to be expected no matter which the initial proportion of Ly$\beta$ and Ly$\gamma$ photons is, as it only depends on the optical depth at the two wavelengths. A spectrum with a larger flux in Ly$\beta$ might produce a \ha/\hb\ ratio larger than the case B one, but the succession between increasing and decreasing stretches would persist.

This interpretation of the \ha/\hb\ behaviour has independently been verified by running models identical in all respects to model M1 except for the stellar continuum, in which small wavelength ranges around the Lyman lines wavelengths have been selectively suppressed. For example, in the model with no continuum between 1000 and 1050 \AA\ the \hahb\ enhancement disappears completely, leaving only a \hahb\ depression around $r\sim$ 10 pc. For brevity, the detailed description of these models has been omitted.

The difference in fluorescent enhancement of different lines is illustrated by Table~\ref{tab:HR2vsNoPumping} for \ha, \hb\ and \hg. These differences are model-specific and depend on the number of photons available to excite the various Lyman lines (which in turn depends on the precise shape of the spectrum and, in a real nebula, on line broadening and Doppler shifts), and on the electron distribution among different sublevels of a given level, which influences the downward radiative paths available and which depends on the local gas density ($l$-mixing is quite ineffective at such low densities). We cannot say whether current synthesis model are accurate enough for these differences to be real; but, if we take them at face value, the straightforward consequence of this differential enhancement is a change in the intrinsic Balmer decrement with respect to its case B value.

Table~\ref{tab:Ha2Hb} quantifies this effect for model M2, by comparing its \hahb\ values to those of model M3. The case B value for these models' temperature is 2.81; the differences between such value and the numbers in the table are determined by the competition among three factors.
On one hand, fluorescent excitation enhances \hb\ more than \ha\ (Table~\ref{tab:HR2vsNoPumping}): this tends to {\it decrease} the Balmer decrement. On the other hand, collisional excitation enhances \ha\ more than \hb\ \citep{L50}, tending to {\it increase} the Balmer decrement. Finally, fluorescent excitation takes place in the center and is therefore favoured in 1D integrations, whereas collisional excitation takes place in the external part of the model and is favoured in 3D integrations. These dependencies explain the patterns seen in the table: in model M2 (with a ``normal'' amount of continuum pumping), fluorescence bluening dominates over collisional reddening in the 1D aperture, while collisional reddening dominates in the 3D aperture; in model M3 (where continuum pumping has been suppressed), fluorescence is quenched, so all the ratios are larger than the case B decrement, and the difference increases proportionally to the weight given to external regions due to the increasing role of collisional excitations.
The consistency of this interpretation has been verified by computing density-bounded versions of models M2 and M3, labelled M4 and M5. In M4 and M5, the nebula is truncated at a radius of 70 pc, so that the region of collisional excitation is not included in the computation: as a result, the \hahb\ ratios are either lower than (M4, density-bounded variant of M2) or close to (M5, density-bounded variant of M3) the case B value.

\subsubsection{Observational implications of gradients in the Balmer decrement}

We can go a bit further by recalling again that the optical depth is different for different Lyman-line photons, in such a way that they are selectively absorbed at different depths into the nebula. Note that this is true no matter which the initial proportion of Lyman-line photons is, so that no role is played here by the caveat of Section~\ref{sec:effect_Balmer_dec} regarding the accuracy of the spectrum. As by-products of the absorption of Lyman photons, different Balmer lines are also produced in different proportions at different depths into the nebula (Figure~\ref{fig:Ha2Hb_HaHb_r}, bottom): as a consequence, the purely fluorescent \hahb\ ratio is expected to vary as a function of depth, being in general different from the case B value (Figure~\ref{fig:Ha2Hb_HaHb_r}, top). Therefore, the projected \hahb\ ratio also depends on the particular direction considered, as it depends on the proportion of fluorescence-excited atoms intercepted along the particular line-of-sight, and therefore on the distance from the exciting stars. Of course, the amplitude and scale length of this variation depend on the scale of the fluorescence zone, which depends on nebular geometry and is probably very thin in most objects; specifically, it will be shown in Section~\ref{sec:nebular_geometry} that the more plane-parallel is the gas distribution, the thinner is the fluorescence zone. But in regions with thicker geometries, such dependence may lie above the spatial resolving power and be misinterpreted as a spatial variation of interstellar extinction. 

\begin{figure}
\plotone{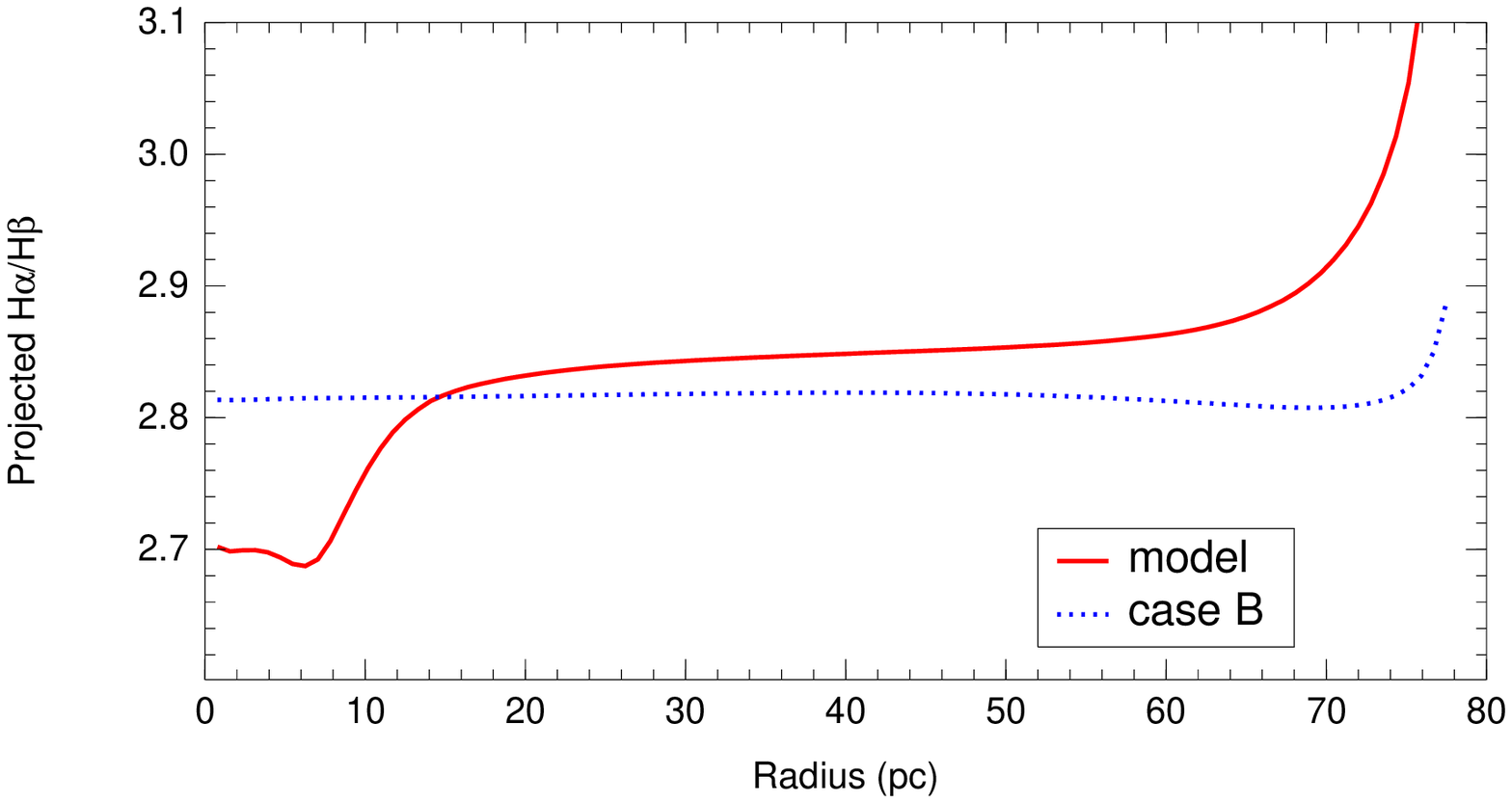}
\caption{Solid line: \hahb\ projected profile along the nebular midplane. Dotted line:  \hahb\ projected profile computed assuming case B emission.\label{fig:Ha2Hb_projected}}
\end{figure}

There are numerous observations reporting systematic changes in the Balmer decrement along different lines of sight and attributing them to differences in dust extinction \citep[see, e.g.,][]{Mal98, Cal02, Kal08}. To investigate the possible role of continuum pumping in such differences,
we simulate spatially resolved observations by plotting the predicted \hahb\ ratio along a simulated narrow slit (Figure~\ref{fig:Ha2Hb_projected}). Note that this is not the same as the 2D integration, which describes the {\sl integrated} emission of a narrow slit.
Figure~\ref{fig:Ha2Hb_projected} shows that, with fluorescent excitation properly taken into account, the \hahb\ ratio actually increases from a minimum value near the center to a maximum in the outskirts. With no other observations available, this pattern could be interpreted as a variation in true extinction and bias the dust distribution inferred from the \hahb\ ratio. 

We believe that, for the most part and in most cases, the observed \hahb\ gradient does indeed reflect the dust distribution; furthermore, it makes physical sense that such distribution is correlated to the spatial position of massive stars. However, we wish to call attention to the fact that a part of the gradient might also be intrinsic, arising as a consequence of the spatial segregation of fluorescent enhancement and collisional excitation of H$^0$. Even adopting a conservative stance, the evidence presented in this section should be interpreted as a measure of the minimum uncertainty affecting interstellar extinction measurements. A rough estimate based on model M2 situates such uncertainty around $\pm$0.02 dex in \hahb, or $\pm$ 0.05 dex in $C($\hb$)$.

\begin{table}
\begin{center}
\caption{Dependence of predicted \hahb\ ratios on fluorescent and collisional excitation.\label{tab:Ha2Hb}}
\begin{tabular}{ccccc}
\tableline\tableline
Integration & \multicolumn{4}{c}{\hahb} \\
\cline{2-5}
  & M2 & M3 & M4 & M5 \\
\tableline
1D & 2.70 & 2.84 & 2.68 & 2.83 \\
2D & 2.82 & 2.85 & 2.80 & 2.83 \\
3D & 2.86 & 2.86 & 2.82 & 2.83 \\
\tableline
\end{tabular}
\end{center}
\end{table}

\section{Influence of the spatial distribution of stars}\label{sec:spatial}

Assuming that the synthesis model used in this work is a good approximation to an actual SED, the effect of continuum pumping is inescapable. On the other hand,
a natural objection to this conclusion is that it would cause apparent abundance gradients of appreciable amplitude, which would have hardly gone unnoticed. Existing evidence is not conclusive: for example, in the works by \citet{Ial06} and \citet{Kal08}, radial gradients in the inferred abundance can neither be confirmed nor dismissed. There may be several causes for this. First of all, the error bar in the derived abundances is of the same order of the effect we are concerned with. Secondly, we are considering here a spherical geometry, whereas there are observational evidences of a blister geometry in some resolved H{\sc~ii} regions; in such a case, the region affected by fluorescence enhancement would be compressed in a thin shell. Finally, the spatial distribution of stars is not always centrally concentrated as in our na\"\i f, highly idealized model. This comment mainly concerns
 giant H{\sc~ii} regions, in which the scale of the system exceeds the Str\"omgren radius of individual stars. In such cases, we can expect a complex geometry, which cannot be represented by our simple model. In the extreme case of homogeneously distributed sources, local predictions would approach the volume-averaged result of a centrally concentrated source.

\section{Dependence of fluorescent intensity on the model's parameters}\label{sec:dependencies}

\subsection{Dependence on stellar population's parameters}\label{sec:stellar_parameters}

In this section we will briefly discuss how fluorescent excitation, whose effectiveness is a function of the Lyman-line photons available in the exciting spectrum, is affected by the stellar population's parameters. Only a qualitative description of the input synthesis models will be used to this aim. The discussion below is based on models with $Z=0.02$ and $Z=0.001$ and ages between 1 and 15 Myr.

The outcome of a comparison between different SEDs depends on which variables are held fixed. In the present case the comparison must be carried out at fixed $Q(\mathrm{H}^0)$, because we wish to compare the Balmer line intensity due to fluorescent excitation to the intensity due to recombination, which is determined by $Q(\mathrm{H}^0)$. 

\subsubsection{Age of the burst}

\begin{figure}
\plotone{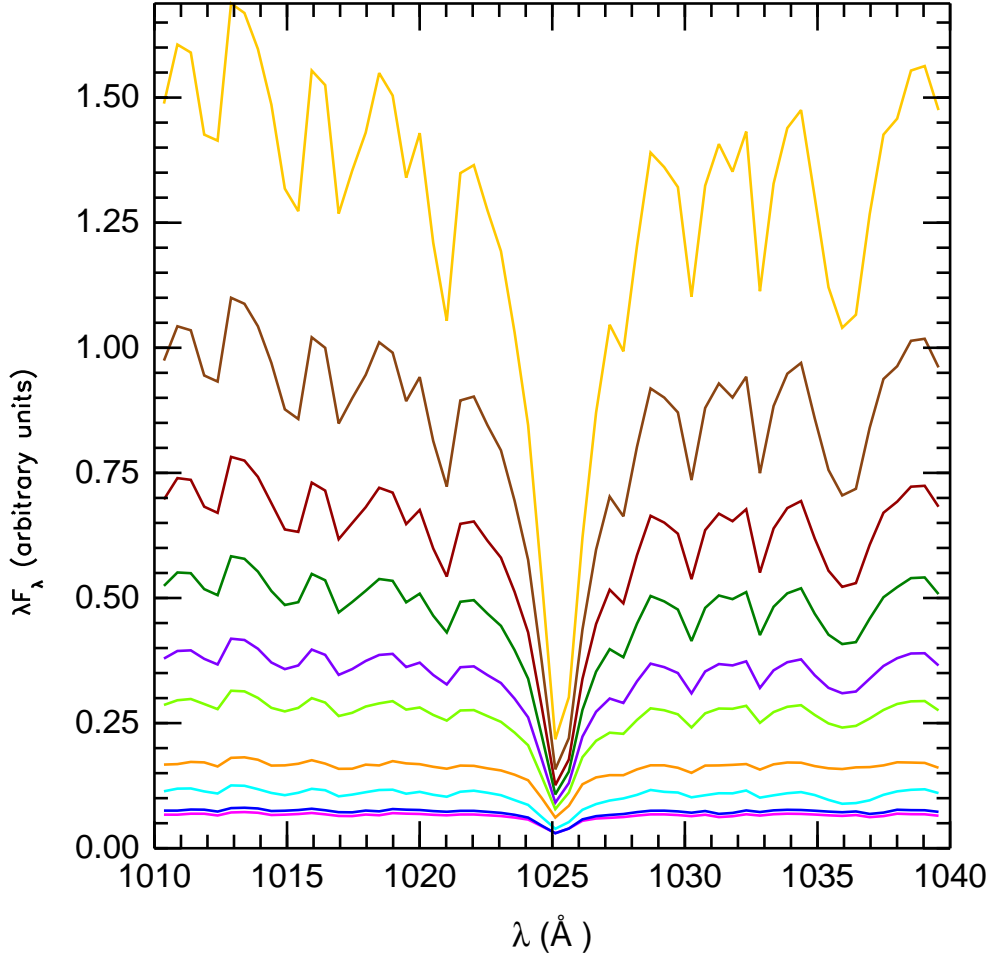}
\caption{SEDs around Ly$\beta$ of a sequence of synthesis models with increasing ages, normalized to the number of ionizing photons. From the bottom up: 2, 1, 3, 4, 5, 6, 7, 8, 9, and 10 Myr.\label{fig:age}}
\end{figure}

The fraction of \ha\ provided by fluorescent excitation monotonously increases as age increases, amounting to, e.g., 3\%, 4\% and 6\% at $t=4$, $6$ and $10$ Myr, respectively. Note that this increase is triggered by the progressive decrease in the number of ionizing photons, which declines with age; this is only partially offset by the increase in the absolute value of the equivalent width of the Lyman $\alpha$ absorption line taking place after 3 Myr. This can be seen in Figure~\ref{fig:age}, which shows a sequence of SEDs in the age range 1-10 Myr normalized to the number of ionizing photons.

\subsubsection{Metallicity}

Comparison of models spanning a wide metallicity range ($Z_{gas}, Z_*=0.001-0.02$) suggests that metallicity has no effect in very young objects, and only a slight effect in older objects, in the sense of larger fluorescent contributions at larger metallicities. Our predictions for the 3D integration of models with ages in the 1-5 Myr range are listed in Table~\ref{tab:Z}. Relative contributions for other apertures might be estimated on the basis of these. 

\begin{table}
\begin{center}
\caption{Fraction of total intensity in H$\alpha$ contributed by fluorescent excitation as a function of metallicity and age.\label{tab:Z}}
\begin{tabular}{lccccccc}
\tableline\tableline
 \multicolumn{1}{c}{$Z$} & \multicolumn{3}{c}{$t$ (Myr)} \\
    & 1 & 3 & 5 \\
\tableline
0.001 & 0.02 & 0.02 & 0.03 \\
0.004 & 0.02 & 0.02 & 0.04 \\
0.008 & 0.02 & 0.03 & 0.04 \\
0.02  & 0.02 & 0.03 & 0.04 \\
\tableline
\end{tabular}
\end{center}
\end{table}

\subsection{Dependence on gas distribution}

\subsubsection{Nebular geometry}\label{sec:nebular_geometry}

By ``nebular geometry'' we exclusively refer to the distinction between thick (spherical) and thin (plane-parallel) models. The volume-integrated fluorescent contribution does not depend on geometry: roughly speaking, the relative excitation rates depend on the number of photons available for each process (ionizing photons in the case of recombination, Lyman-line photons in the case of fluorescent excitation). On the other hand, the beam-integrated fluorescent contribution does depend on nebular geometry: the more plane-parallel the model, the closer its beam- and volume-integrated line intensities. Asymptotically, the intensities seen through different integrations coincide.

\subsubsection{Nebular boundedness}

Density-bounded models differ from ionization bounded models in that the outermost part is missing.
The fraction of fluorescent excitation in these models increases proportionally to the extension of the missing part, until the slab becomes so thin that it falls under case C.

\subsection{The effect of dust}\label{sec:dust}

Our simulations so far have assumed a dustless nebula. This is not a realistic assumption in most nebulae, and the difference may be relevant in the context of this research since dust grains absorb UV photons. To understand this issue, we have computed a series of dusty models with varying metal abundances and dust content. Our models, which span the metallicity range $0.004 \le Z_{gas} \le 0.016 $, show no dependence of the relative fluorescent contribution on dust content (although, of course, absolute fluxes are altered). The reason for this behaviour is the approximate constancy of dust opacity in the relevant range, which roughly goes from the energy of Ly$\beta$ to a few Rydbergs. 
The grains used in these models are Orion-type, that is relatively large; using ISM-type grains, with a distribution extending to smaller wavelengths, UV photons would be more effectively blocked, resulting in smaller and dimmer nebulae. However, the fluorescent relative contribution would stay constant. 
This is a consequence of the small relative variation of dust opacity in the relevant range, which goes from Ly$\beta$ to $\lambda\sim 400$ \AA\ (this range encloses more than 90\% of the total number of ionizing photons); see, e.g., Figure~19 of \citet{Bal91}, which shows that the total variation of the absorption coefficient in the relevant range is within a factor of two, which defines the maximum bias that might affect our dustless results.

\subsection{The effect of non-thermal line widths}\label{sec:turbulence}

The line widths of actual H{\sc~ii} regions are larger than the expected thermal width. The origin of this non-thermal contribution is currently unknown, although it has been suggested that it might be turbulence \citep{Ral98, Ral05}.
Whatever its cause, the effect of an increased line width is to make continuum pumping more efficient by increasing the line flux available \citep[see, e.g.,][]{F92}. Since our models so far have been computed under the assumption of thermal line width, we will discuss here the consequences of assuming increasing line widths, assuming they are caused by turbulence. 

\begin{figure}
\plotone{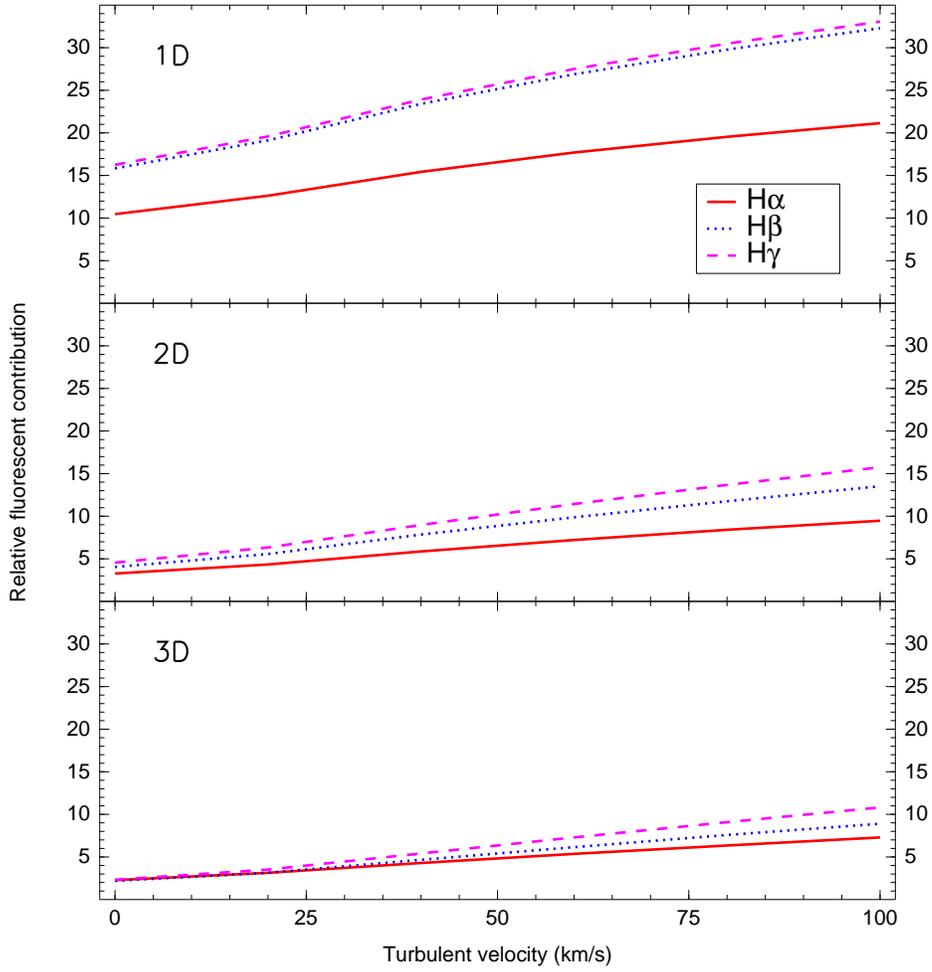}
\caption{Relative contribution of fluorescent excitation to the intensities in \ha, \hb\ and \hg as a function of microturbulent velocity.\label{fig:turbulence}}
\end{figure}

Figure~\ref{fig:turbulence} illustrates the effect of turbulence on our reference model M2 at different values of the microturbulent velocity, which is assumed Gaussian. The fluorescent component increases linearly with the turbulent velocity in all the range explored; specifically, from 0 to 100 km s$^{-1}$ it doubles in 1D integrations and triples in 2D integrations, whereas the increase in 3D integrations is even larger. This gives an indication of how this paper's estimates should be increased to take turbulence into account. At even larger velocities, the effect would continue to increase.

\section{Dependence of H fluorescent excitation on the stellar atmosphere details}\label{sec:atmospheres}

In this section we will consider the case of H{\sc~ii} regions around individual stars to discuss how our results depend on the details of the stellar atmosphere. 
In Section~\ref{sec:caseC} we mentioned two stellar spectral features, not present in  blackbodies, 
that may affect the balance between the Lyman-line photons and the 
Lyman-continuum photons, and, hence, the relative importance of fluorescent enhancement in Balmer lines: the Lyman jump and the stellar absorption lines. These two features affect continuum pumping in 
opposite directions.
On one hand, the ratio of non-ionizing to ionizing photons depends on the size of the Lyman jump: hydrogen continuum pumping has a greater relative importance in a stellar atmosphere with a Lyman jump than in a blackbody case. On the other,
absorption lines at the Lyman frequencies decrease the number of continuum photons available to pump, thus quenching continuum pumping. Both features depend on the characteristics of the star, i.e. on its
spectral type and luminosity class, metallicity, and strength of the stellar wind;
for example, the size of the Lyman jump is larger in an O9\,V star than in an O3\,V
star.

The stellar flux emitted by massive stars in the spectral range between the Lyman jump 
and 1200 \AA\ has never been studied in terms of its effect on the surrounding 
nebular material: usually, only the ionizing part of the stellar fluxes is considered to be
important. However, the arguments of Section~\ref{sec:mechanism} show the importance of the non-ionizing part.
Fortunately, our knowledge of the characteristics of the stellar flux in this spectral 
range has enormously improved in the last decades thanks to the far-UV spectroscopic 
observations provided by space telescopes \citep[mainly Copernicus and FUSE: see, e.g.,][]{Wal02,Pal02} 
and the development of stellar atmosphere codes for massive stars \citep[e.g.,][]{HM98, Pal01}. We review these
characteristics here and give examples of how they may affect the behavior of photoionization models including continuum pumping.

To this aim we considered two sources of information: the far-UV spectral atlases compiled
by \citet{Wal02} and \citet{Pal02}, and the predictions of WM-basic models \citep{Pal01}. Below, we describe the main stellar spectral features that
may affect case D predictions. In particular, we are interested in the Lyman 
jump and those spectral features affecting the stellar flux emitted at the Lyman wavelengths.

\begin{figure}
\plotone{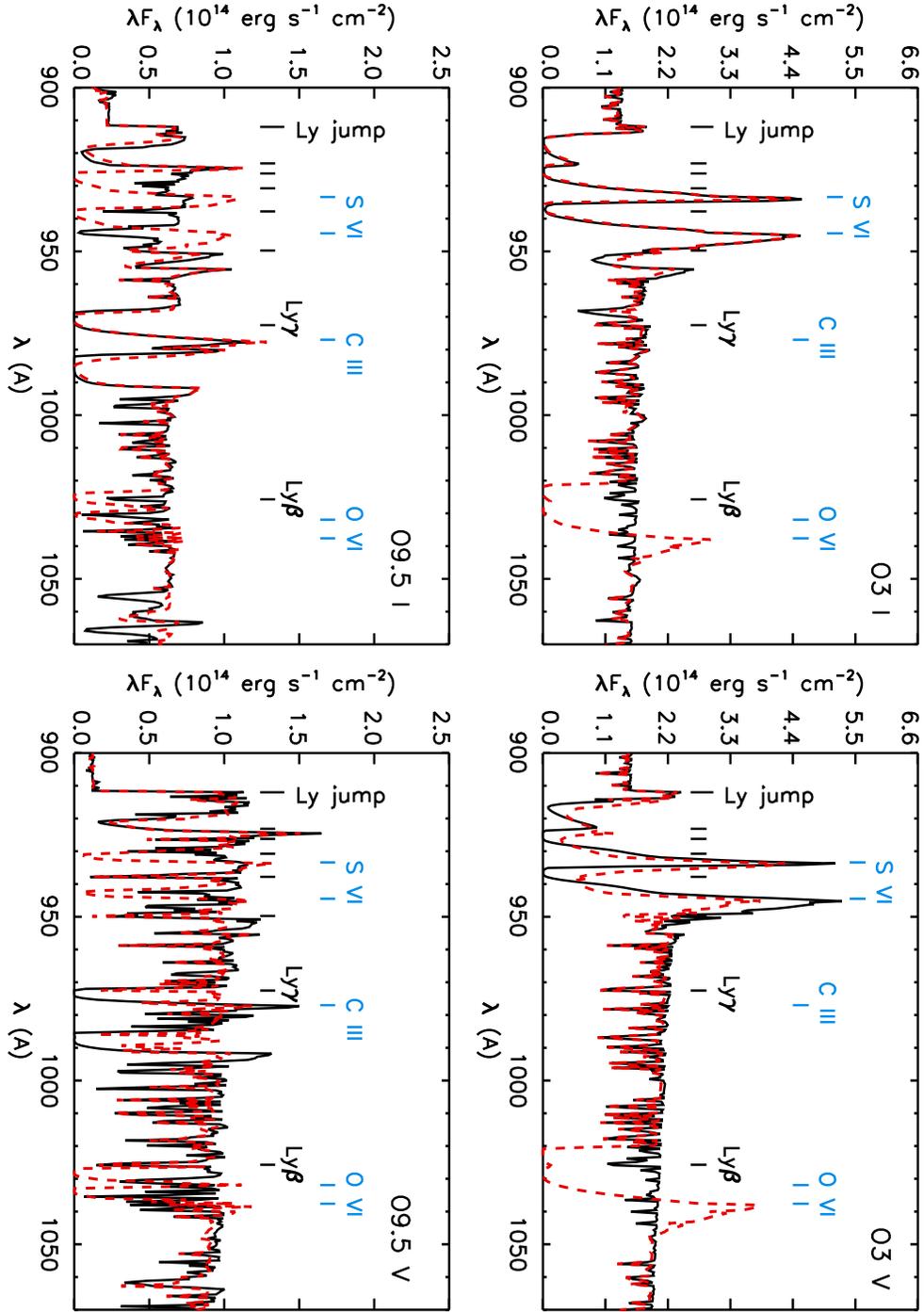}
\caption{WM-basic models for stars of different spectral type and
luminosity class (see \ref{tab:WMbasic_pars} for the main models' properties). 
Black solid lines: models with a smooth wind velocity law; light dashed 
lines: shocked-wind models. The main stellar spectral features 
that may affect case D predictions are indicated.\label{fig:WMbasic_SpT}}
\end{figure}

In Figure~\ref{fig:WMbasic_SpT} we show the spectra of
WM-basic models corresponding to four different spectral type-luminosity class combinations, in the wavelength range 900 - 1070 \AA, which includes the Lyman jump, the relevant H Lyman lines and other stellar features of interest.
Only this range is represented because we are only interested in those lines that can produce Balmer lines through continuum pumping, i.e. Ly$\beta$ and Lyman lines of shorter wavelengths. Two types of
WM-basic models were computed and included in the figure (see also Table~\ref{tab:WMbasic_pars}): black solid lines represent models with a smooth wind velocity law, 
light dashed lines represent models including the effect of shocks in the wind. 
This second type of models are necessary to reproduce some of 
the lines observed in the far-UV spectra of massive stars, such
as O{\sc~vi} lines (see below).

\subsection{The Lyman jump}

This feature, located at $\lambda = 911$ \AA, is the limit between the ionizing and the non-ionizing 
part of the stellar flux. It is produced by the presence of a sharp limit in the ionization cross section for the ground level of H{\sc~i}. Therefore, its size 
depends on the amount of neutral H{\sc~i} atoms in the atmosphere, which is determined by the temperature and density in the region of the stellar atmosphere 
where the stellar continuum at this wavelength is formed. Generally speaking, the Lyman jump will be larger at lower
effective temperatures and larger gravities (since in these cases hydrogen is expected
to be more recombined). Note, however, that the size of
the Lyman jump may also be affected by other features of massive stars' winds \citep[see, e.g.,][]{Gal89,Nal96}.

Since practically no observations of the Lyman jump exist (the non-ionizing part of
the stellar flux is generally not accesible to direct observations), we have to rely on
stellar atmosphere models. As already mentioned,
the Lyman jump is larger in stars with larger gravities (dwarf stars) and later spectral types (Figure~\ref{fig:WMbasic_SpT}).

\subsection{Stellar lines}

Both observations and models show that the stellar flux at the Lyman wavelengths may be affected not only by the H Lyman lines, but also by 
resonant lines of heavy elements at different wavelengths that may develop P-Cygni profiles if the
characteristics of the stellar wind are appropriate. If the
P-Cygni profile is completely developed and the wind terminal velocity is large
enough, the saturated blueshifted absorption may reach bluewards the wavelength of
the closest H Lyman line (see Figure~\ref{fig:WMbasic_SpT}). 

As illustrative examples, we discuss here the behavior of the O{\sc~vi} $\lambda\lambda$\,1032, 1038  doublet and the C{\sc~iii} $\lambda$979 lines as a function of spectral type, luminosity class and wind characteristics.
Figure~\ref{fig:WMbasic_SpT} also shows the location of the S{\sc~vi} lines, which may affect other H Lyman lines.
However, this is only given for information and its behavior is not discussed here.

The O{\sc~vi} $\lambda\lambda$\,1032, 1038 lines may affect the spectral
region of Ly$\beta$. The spectroscopic catalogues by \citet{Pal02} and \citet{Wal02}
show that these lines exhibit a single P-Cygni wind profile in early-type O stars, both in dwarfs and supergiants. In later O-type spectra, two distinct blueshifted absorption features are often visible and, in the case of supergiants, O{\sc~vi} lines 
are still visible in B1\,I stars. These lines depend only weakly on the temperature, luminosity, and metallicity of the star, and can still be seen even at metallicities as low as in the SMC.

The presence of lines of this ``superionization'' species in the spectrum
of OB stars has been commonly attributed to the effect of shock-produced X-rays on
the ionization distribution in hot stars winds \citep[see, e.g.,][for a discussion concerning the origin of this exotic species]{Mal93}. In fact,
only stellar atmosphere models including
shocks in the stellar wind can produce the O{\sc~vi} lines. 

Figure~\ref{fig:WMbasic_SpT} shows the synthetic UV spectra of WM-basic models with and without shocks. 
In the case of O3 stars, strong O{\sc~vi} P-Cygni profiles are only developed if shocks are
included in the models. Something similar occurs in the O9.5 models where, although
a P-Cygni profile is not developed, we can still find strong
extended blueshifted absorptions. In all the cases, the presence of the O{\sc~vi} lines
totally quenches the stellar continuum flux in Ly$\beta$. This only
happens if the terminal velocity of the wind is above $\sim$1700 km\,s$^-1$. 

Our second example is the C{\sc~iii} $\lambda$979 line. \citet{Pal02}
show that, in supergiants, this line appears as a broad P-Cygni profile with a broad absorption component
in spectral types from O6 to 
O9.7, while a broad absorption feature is detectable in subtypes O9.5 
to B3 V. On the other hand, \citet{Wal02} showed that in SMC supergiants 
the line saturates despite the lower abundances, and does not appear weaker than in 
their LMC counterparts. 
Figure~\ref{fig:WMbasic_SpT} shows the behavior of this line as predicted by WM-basic models 
for the stellar cases considered. In both O9.5 models the C{\sc~iii} line develops
a P-cygni profile whose blue absorption affects the flux in Ly$\gamma$.
Interestingly, in the O9.5V star this line is inhibited by the inclusion of shocks.

\subsection{Implications of the non-ionizing stellar SED on case D predictions}

In this section we discuss how the spectral features considered 
above make ionized nebulae deviate from case B to case D.
To this aim we constructed a set of spherical, dustless, constant density
photoionization models with Cloudy. The internal radius and density 
of the models were $R_{\rm i}$\,=\,1 pc and 
$n_{\rm H}$\,=\,200 cm$^{-3}$, respectively, and no turbulent line widening was assumed.
The SEDs predicted by WM-basic for 
the stellar models summarized in Table~\ref{tab:WMbasic_pars} were 
included as input in these Cloudy models, normalized to 
log\,$Q$(H$^0$)\,=\,52.09 s$^{-1}$.

\begin{figure}
\epsscale{1.0}
\plotone{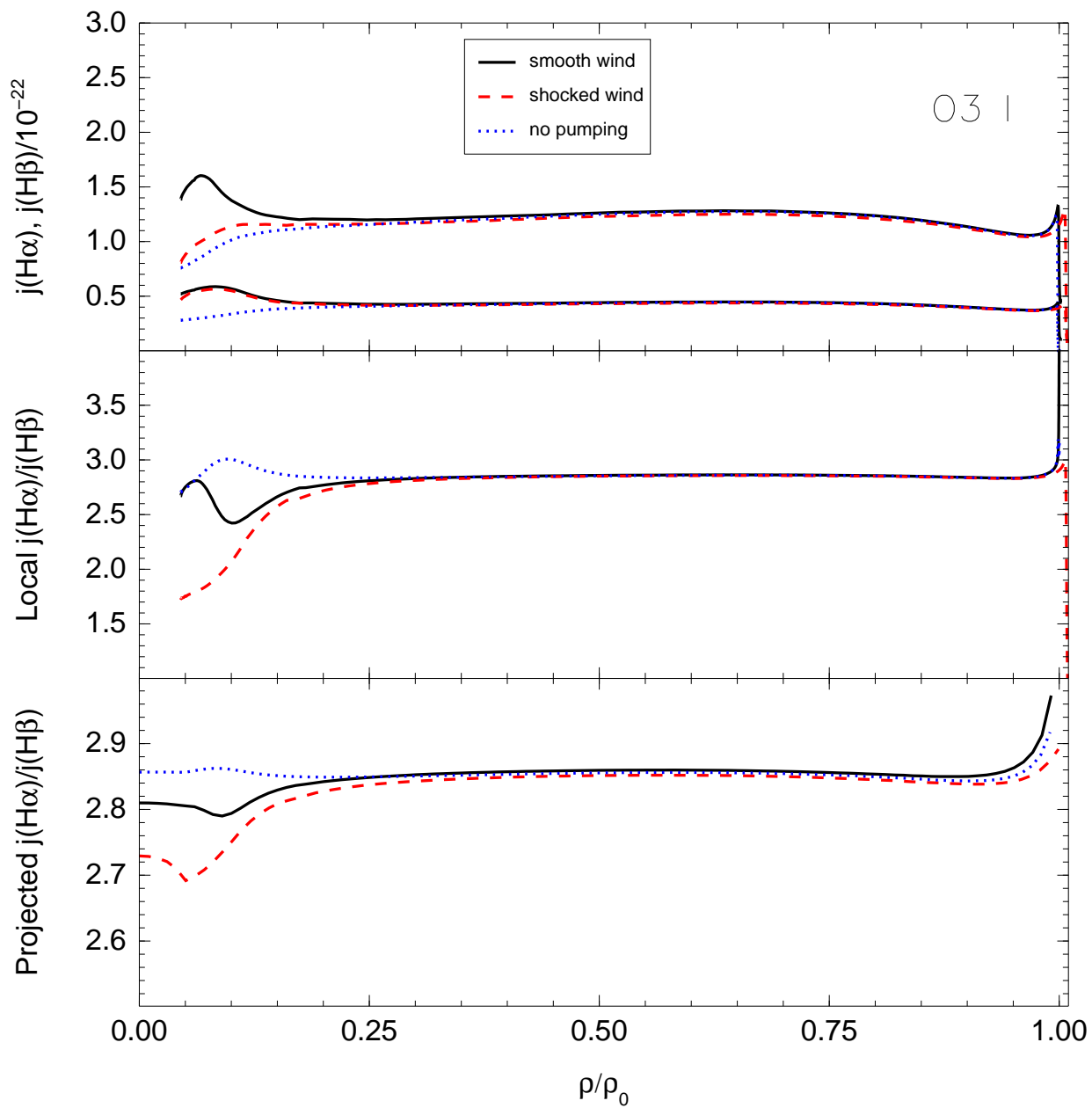}
\caption{Local emissivities, local Balmer line ratios and projected Balmer line ratios of photoionization models 
illuminated by WM-basic models of an O3 I star,
with a smooth wind velocity law (solid) and a shocked wind (dashed), 
compared to those of a model with no fluorescent excitation (dotted). Emissivities 
are in units of erg s$^{-1}$ cm$^{-2}$. The distances to the stars are normalized to the maximum 
nebular radius in each case.\label{fig:o3i}}
\end{figure}

\begin{figure}
\epsscale{1.0}
\plotone{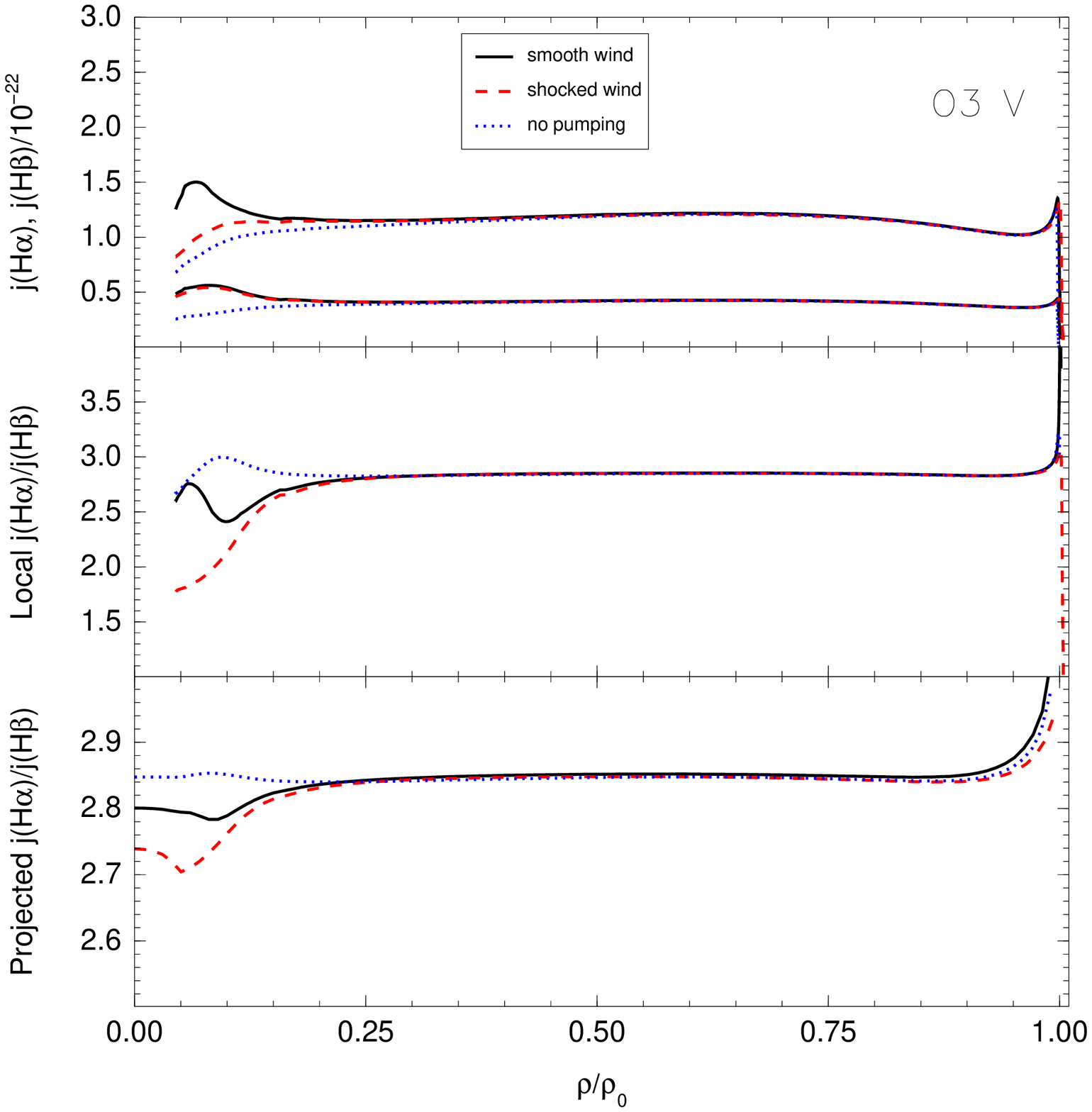}
\caption{As in Figure~\ref{fig:o3i}, for the case of an O3 V star.\label{fig:o3v}}
\end{figure}

\begin{figure}
\epsscale{1.0}
\plotone{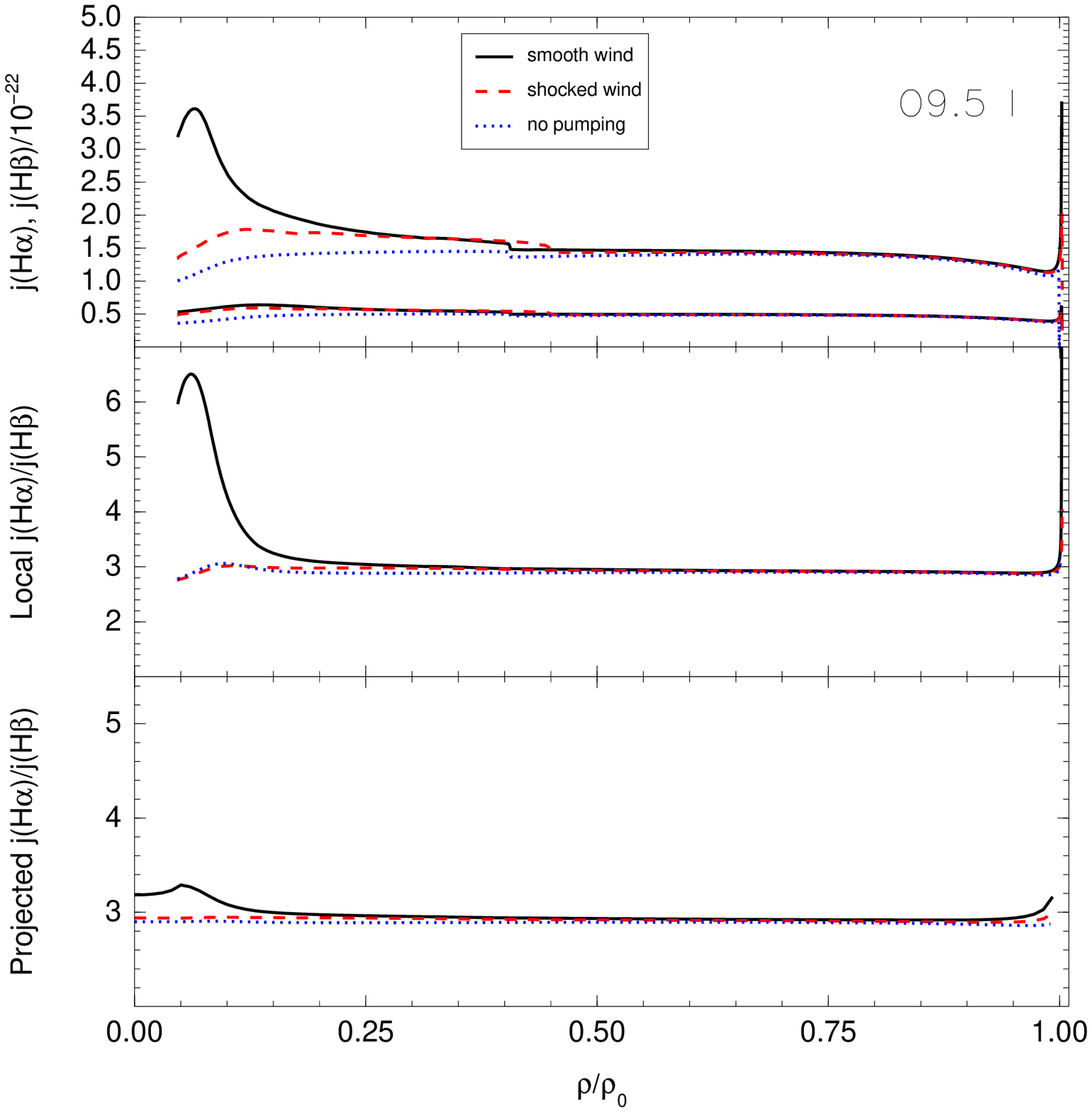}
\caption{As in Figure~\ref{fig:o3i}, for the case of an O9.5 I star.\label{fig:o95i}}
\end{figure}

\begin{figure}
\epsscale{1.0}
\plotone{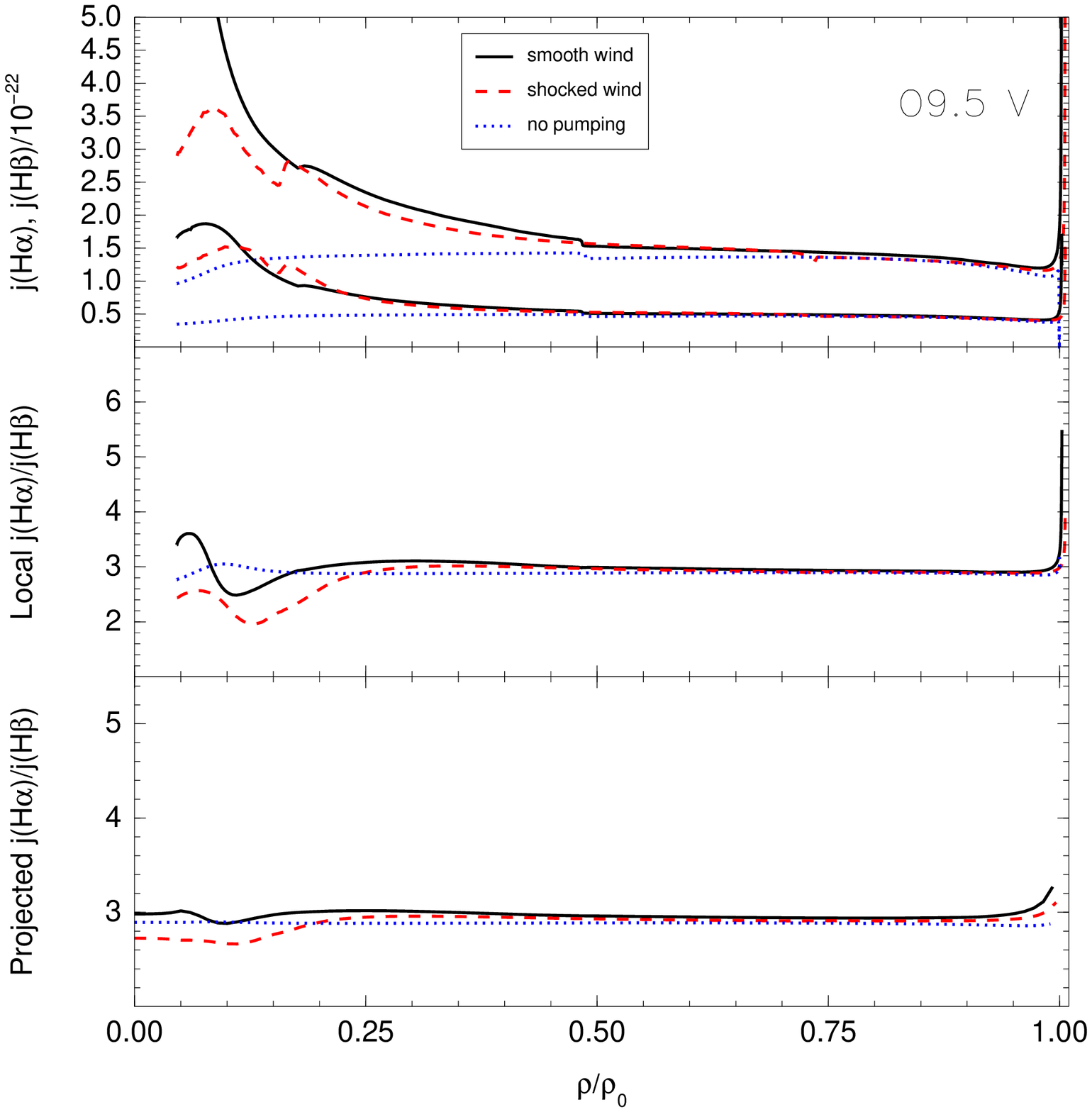}
\caption{As in Figure~\ref{fig:o3i}, for the case of an O9.5 V star.\label{fig:o95v}}
\end{figure}

Figures~\ref{fig:o3i} through~\ref{fig:o95v} 
show the predicted  
H$\alpha$ and H$\beta$ emissivities and the \hahb\ 
ratio for different stellar cases: similarly to 
Figure~\ref{fig:WMbasic_SpT}, dashed and solid lines indicate results 
obtained with or without shocks, respectively, 
with full inclusion of fluorescent excitation, while dotted lines 
are results obtained by switching fluorescent 
excitation off. The behavior of ratios can be easily understood by considering 
the spectral features of each model as a function of its properties.

In O3 stars, we see that while the behavior of the \hahb\ line ratio 
is very close to case B predictions when a smooth stellar wind velocity law is 
assumed, it is decreased with respect to
case B predictions when a shocked wind stellar model is considered.
In the case of a smooth wind law, even though the Lyman jump is small and the stellar
continuum flux at the Ly$\beta$ and Ly$\gamma$ wavelengths is dimished, 
the local
H$\alpha$ and H$\beta$ emissivities 
are noticeably affected by stellar continuum pumping
in the region close to the stars, resulting in 
a strong deviation from case B predictions. However, the H$\alpha$ and H$\beta$ 
variation are similar, in such a way that their ratio is
close to case B predictions.
On the other hand, in the shocked-wind stellar atmosphere models, the presence 
of the O VI P-Cygni line (see Figure~\ref{fig:WMbasic_SpT}) leaves no Ly$\beta$ photons, and hence 
the fluorescent contribution to H$\alpha$ is negligible. Since the 
fluorescent excitation of H$\beta$ is similar to the previous case, the combined
effect is a lower \hahb\ ratio in the inner part of the nebula 
as compared to case B predictions.

A similar line of reasoning can be followed to understand the other stellar 
cases represented in Figures~\ref{fig:o3i} through~\ref{fig:o95v}.
For example, the large H$\alpha$/H$\beta$ line
ratio predicted in the O9.5\,I case when a smooth wind law is used in
the stellar atmosphere model calculation is a consequence of the large
flux in Ly$\beta$ (which triggers a large fluorescent contribution
to \ha) and the small flux in Ly$\gamma$, due to the presence of the C\,III 
line (which quenches the fluorescent contribution to \hb). (Note that in this case
the Lyman jump is larger.) This large deviation from case B vanishes
when the shocked-wind stellar atmosphere model is considered, due to the
development of a broad absorption feature in the O{\sc~vi} lines that affects Ly$\beta$.
In this case, the continuum pumping at Ly$\beta$ and Ly$\gamma$ is zero (note,
however, that the H$\alpha$ and H$\beta$ emissivities slightly deviate from
case B. This is produced by pumping in Ly$\delta$).

This example illustrates how deviation from case B to case D depends
on the spectral features appearing in the spectral region considered. 
Generally speaking, we can conclude that the characteristics of the wind has a sizeable effect 
on the predicted Balmer line ratios. Models with shocked winds have lower \hahb\ 
wih respect to models with smooth wind-velocity laws. 

\begin{table}
\begin{center}
\caption{Parameters of the WM-basic models used in this work.
For those models including a shocked wind, the X-ray luminosity is
$L_{\rm X}$ = 10$^{-6.5}$ $L_{\rm bol}$. The parameters are based on the 
observational $T_\mathrm{eff}$ scales by \citet{Mal05b} (their Table 4).
\label{tab:WMbasic_pars}}
\begin{tabular}{lrrrrr}
\tableline\tableline
& & \\
\multicolumn{1}{c}{SpT}   &  \multicolumn{1}{c}{$T_{\rm eff}$} & \multicolumn{1}{c}{log\,$g$} &  \multicolumn{1}{c}{$R$} & \multicolumn{1}{c}{log\,$\dot{M}$}  & \multicolumn{1}{c}{$v_{\infty}$} \\   
 & \multicolumn{1}{c}{($K$)} & \multicolumn{1}{c}{(cm s$^{-2}$)}  &  \multicolumn{1}{c}{($R_{\sun}$)} & \multicolumn{1}{c}{(M$_{\sun}$\,yr$^{-1}$)} & \multicolumn{1}{c}{($km\,s^{-1}$)} \\   
& & \\
\tableline
O3$\,$I   & 42233 & 3.73 & 18.6 & -5.15 & 2800 \\
O9.5$\,$I & 30463 & 3.19 & 22.1 & -5.67 & 2200 \\

O3$\,$V   & 44852 & 3.92 & 13.8 & -5.77 & 2800 \\
O9.5$\,$V & 31884 & 3.92 &  7.2 & -6.89 & 1300 \\
\tableline
\end{tabular}
\end{center}
\end{table}

\subsection{The effect of the Lyman line profile}\label{sec:line_profile}

The discussion above has made clear that the amount of Lyman continuum pumping in a nebula is crucially dependent on the line profiles.
These depend on both the wind structure and the relative motion between the stellar source and the gas, so 
the case should be considered in which the Lyman lines in the stellar continuum are in emission
rather than absorption; in such a case, the nebula could produce more excitation than expected.
We have taken such possibility into account by computing a sequence of models with varying blocking factors, whose effect is to amplify the amount of continuum pumping by a given factor, which can be either smaller or larger than one \citep[see Appendix A in][]{Pal07}.

\begin{figure}
\plotone{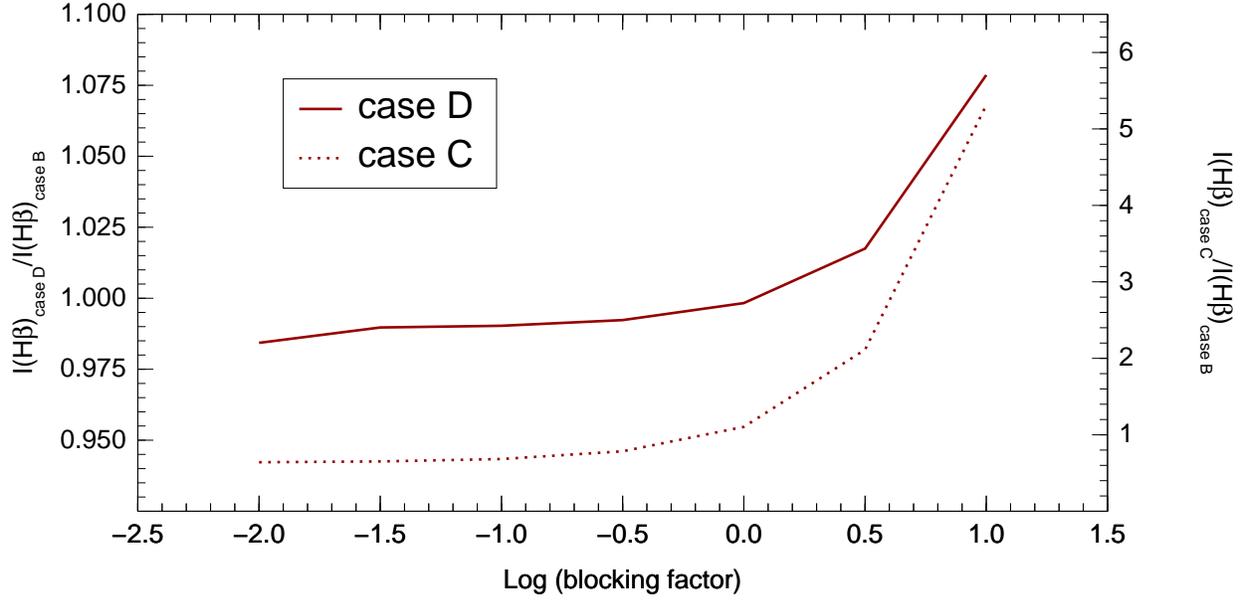}
\caption{$\rm{H}\beta_{\rm tot}/\rm{H}\beta_{\rm case B}$ as a function of blocking factor, for a model nebula ionized by a stellar model with a smooth wind law. Solid line, left-hand scale: radiation-bounded model (case B conditions). Dotted line, right-hand scale: predictions for the innermost zone (case C conditions). Note that the ratio is, in case C, much larger than 1. See also \citet{F99}.\label{fig:bf}}
\end{figure}

The result is illustrated by Figure~\ref{fig:bf}, which shows the ratio between the total H$\beta$ emission and the case B H$\beta$ emission as a function of the blocking factor in a radiation-bounded spherical model ionized by a stellar continuum (solid line, left-hand scale). On the same figure, we have plotted the predictions for the first zone of the same models, a case corresponding to classical case C. In either case, a blocking factor larger than 1 implies that pumping is amplified with respect to the predicted value. It can be seen that increasing values of pumping trigger \hb\ intensities well above predictions, particularly under case C conditions.

\section{Discussion}\label{sec:discussion}

\subsection{Comparison with collisional excitation}

In this section we highlight the differences between fluorescent and collisional excitation, in order to provide some useful guidelines to minimize their impact on observations aimed at abundance determinations. Both mechanisms provide channels to increase the intensity of Balmer lines, but they act with different effectiveness depending on the location within the nebula and the physical conditions regime. The following is a quick summary of such analogies and differences.

\subsubsection{Why fluorescent excitation might be more important than collisional excitation}
Along any given line of sight, fluorescent excitation takes place in the nebular region which lies the closest to the illuminating stars. Collisional excitation, on the contrary, takes place preferently in the recombination zone that marks the outer limit of the H{\sc~ii} region. Such region may be lacking totally or in part if the nebula is density-bounded, whereas a fluorescent region always exists.
A further consequence of the different position in the nebula is that the relative contribution of fluorescence is enhanced in slit observations.
Finally, fluorescence appears to be happening in all kind of nebulae, with no strong preferences for metallicities and with a tendency to increase at increasing ages, whereas collisional excitation is effective only in hot, young nebulae. All in all, then, fluorescence appears to be a more ubiquitous process.

\subsubsection{Comparison between the uncertainties affecting either process}

The {\sl integrated} value of fluorescent excitation does not depend on the geometrical distribution of the gas; neither does the {\sl integrated} value of collisional excitation, provided the region is radiation bounded. Therefore, the two processes are pretty much geometry-free as long as the whole nebula is integrated, and a detailed knowledge of geometry is not necessary.

Fluorescent excitation might depend much more than collisional excitation on the distribution of stars in the case of spatially resolved observations. This circumstance is further complicated by the fact that it is not necessarily dominated by the hottest stars, as are recombination and, probably, collisional excitation.

Both processes also depend on the illuminating SED, but collisional excitation depends on it in a much weaker way than fluorescent excitation. The dependence of collisional excitation on the SED arises on the heating power of the ionizing photons: the harder the average photon, the hotter the gas, so younger clusters are more collisionally excited than older clusters. However, a rough knowledge of the cluster's age and its SED is usually sufficient to this aim. On the other hand, fluorescent excitation crucially depends on the SED, and specifically on the detailed profile of the relevant spectral features, which determine the proportion of photons available to excite the various Lyman lines. Theoretically, such shape depends on the stellar mix, the metallicity and details of the atmosphere. Examples of this are given in Section~\ref{sec:atmospheres} and in \citet{GDal97}, where it is shown how the stellar Ly$\beta$ flux may be affected by a neighbouring O{\sc~vi} line. 
This extreme dependence implies that moderate nebular motions might drastically change the result by Doppler-shifting the line photons seen by the gas. In sum, uncertainties related to the SED affect the estimation of fluorescent excitation, while they have a much smaller impact on the evaluation of collisional excitation.

As for the \hahb\ ratio, it can only be increased by collisions, while it can either be increased or decreased by fluorescence, depending on the details of the SED and the spatial position considered.

\subsection{Literature on case D}

As noticed in the Introduction, not much has been written on the subject of optically thick nebulae illuminated by stellar radiation, which we have called ``case D''. There are several specific reasons for this, at first surprising, fact. One of them has already been mentioned earlier in this section: absorption lines in stellar continua are expected to quench continuum pumping. Even if the hypothesis is made that enough photons survive, a quantitative analysis must by necessity rely on a detailed model spectrum; if this ingredient is lacking, even qualitative assessments become difficult to make (see, e.g., Section~\ref{sec:caseC}). Only recently have high-resolution stellar libraries and synthesis spectra become available, making possible this kind of study. From this point of view, the forthcoming publication of high-resolution spectral synthesis models will be a giant leap forward.

For lines of other ions, fluorescence is a well-known excitation mechanism \citep[see, e.g.,][]{S68,F92,L95}. Its effect on the helium spectrum has been recently discussed by \citet{BSS02}.
In the case of hydrogen, though, fluorescence only provides a small fraction of the total intensity, and until recently neither theory nor observations were accurate enough to justify an analysis at that level of precision. As an extreme example, consider that at the time of publication of the Baker and collaborators' series of papers, Gaunt factors equal to one were still considered a reasonable approximation; and in 1959, errors on Gaunt factors as great as 20\% were still acceptable (Seaton 1959).

A final point is brought about by improvements in technology, and in particular by the increasingly better spatial resolution.  Since fluorescent contribution is the largest in beam-like observations close to the stars, the availability of spatially-resolved information might actually signify a worsening in accuracy. This is analogous to what happens in population synthesis, where the straightforward modelization of stellar populations is made difficult by the poor sampling associated with small observing elements, in such a way that a probabilistic treatment becomes imperative \citep[i.e.,][]{CL06}.

\section{Summary and conclusions}\label{sec:summary}

Case D is the theoretical scenario describing optically thick nebulae irradiated by stellar continua. In these objects, the Balmer spectrum has a fluorescent component superposed to the recombination component: while the ionizing part of the stellar continuum is responsible for the recombination component, the non-ionizing part produces a generally small but non negligible fluorescent component. In the same way as collisional excitation of $\mathrm{H^0}$, which is discussed in various papers cited in this work, this mechanism represents an additive correction to recombination, and as such it must be taken into account in abundance analysis to avoid biased results. 

Case D represents a correction to the more famous case B, but, as case B, it is a limiting scenario which only asymptotically describes the behavior of real nebulae. The label ``case D'' has been given for symmetry with case C, but cases C and D are inherently different from cases A and B in that they are not descriptions of intrinsic gas properties, but rather descriptions of complex systems which depend as much on the intrinsic properties of hydrogen as on the ``boundary conditions'' represented by the stellar pumping.

To estimate the fluorescent contribution predicted under case D it is necessary to understand how it depends on the object's physical parameters and how it is affected by the observational conditions. With respect to the first point, our bottom line in this work is that a proper modelization requires inputting high-resolution synthesis models into the photoionization code. Even this might be only a rough approximation as Doppler shifts and various line-broadening mechanisms might be acting in real nebulae, thus altering the pattern of photons absorption; but at least it is a physically-based approximation, and a good starting point for further explorations. 

Regarding the influence of observational conditions, we have shown that the relative fluorescent intensity observed depends on the aperture. Such dependence is straightforward in the case of a centrally concentrated cluster, but it becomes non trivial as more complicated spatial configurations are considered. 3D models including continuum transfer and photoionized by high-resolution SEDs might help in these cases, but even these could be insufficient to account for actual complexities, as a substantial fraction of the fluorescent component might come from stars of stellar type later than O. When large apertures are considered, geometrical features bear no influence.

A comparison with the other major deviation from pure recombination, collisional excitation, is interesting both to clarify both mechanisms and to devise observational strategies to deal with them, which usually means trying to minimize them. Our study suggests that fluorescence might be much more important than collisional excitation in two importanty ways: on one hand, it may contribute a larger fraction of the intensity; on the other hand, it is predicted to affect virtually any H{\sc~ii} region, whereas collisional excitation is effective only in hot, young, metal poor, radiation-bounded H{\sc~ii} regions. One important difference between the two is that while collisional excitation takes place in the outer parts of H{\sc~ii} regions, fluorescent excitation takes place in the vicinity of stars: the observer willing to minimize its contribution on the total intensity is thus placed before a dilemma, as the former mechanism is minimized by spatially-resolved observations, while the latter is minimized by observations of the complete nebula. 
Finally, collisional excitations always produce a reddening in the spectrum, 
whereas fluorescence might produce both a bluening and a reddening at different points in the nebula. Both mechanisms must be taken into account when optical spectra are used to study the dust distribution of observed objects, as they can introduce small biases in spatially-resolved extinction measurements.

Since this is an exploratory paper, we do not attempt to establish here an exhaustive calibration of fluorescent excitation of Balmer lines across the whole H{\sc~ii} regions' taxonomy; rather, we aim at pointing out a problem and the tools necessary to solve it. Fluorescent excitation adds a small but non-negligible contribution to the observed Balmer spectra of H{\sc~ii} regions and, as such, it should be studied in detail and thoroughly understood since neglecting it might introduce biases in abundance determinations. Such biases might be particularly significant when determining the abundance of primordial helium, where a high accuracy is required. For such application, it is also necessary to take into account the effect of continuum fluorescence on the helium emission lines.

\acknowledgments
This work has been funded with support from the Spanish Ministerio de 
Educaci\'on y Ciencia through grants AYA2004-63030 and 
AYA2007-64712, co-financed with FEDER funds.
GJF acknowledges support from NSF (AST 0607028) and NASA (07-ATFP07-0124). 
SSD acknowledges support by the Spanish MEC/Fullbright postdoctoral fellowship program.
The title was inspired by L. Lucy's paper. The plots have been made with the IDL package {\it jmaplot} by Jes\'us Ma\'\i z. Fruitful discussions have been conducted with Jorge Garc\'\i a Rojas, Antonio Peimbert, Manuel Peimbert, Enrique P\'erez, Grazyna Stasinska, and Ryszard Szczerba.
VL acknowledges the Instituto de Astronom\'\i a-UNAM for hospitality.

\appendix

\section{Case A, B, and C in C08}\label{sec:appendix}

Version C08 of Cloudy has the ability to resolve atoms and ions of the H-like iso-electronic sequence. For most of its history the code has used the compact model of the H-atom originally described by \citet{F80}, \citet{CF88} and updated by \citet{FF97}. That model assumed a well $l$-mixed $n$-configuration as pioneered by \citet{S59}. Under this assumption the populations of $l$-terms within an $n$ configuration are proportional to the statistical weight 2($l$+1). Term-averaged transition probabilities, such as those given by NIST, are used.

With the advent of faster processors it is now possible to resolve the $nl$ terms within an $n$ configuration. The number of states that must be considered is substantially larger. For instance there are 210 terms within the first 20 levels of H$^0$, so on most processors the linear algebra will be roughly 2 dex slower. In the current implementation it is possible to adjust the number of $n$-configurations which are resulted into $l$-terms. These are then supplemented with several additional $l$-mixed $n$ configurations to ``top off'' the model atom. Only lines produced within the $nl$ resolved levels are reported.

The predictions of the resolved and $l$-mixed models can be different at low densities where the real $l$-terms are not mixed by collisions. Continuum fluorescence, the topic of this paper, has different effects in these two approximations, as stressed by Ferland (1999). The reason is that $l$-averaged transition probabilities between excited $n$-configurations and the ground $n$=1 configuration are dramatically different because only $np - 1s$ transitions occur. At low densities the rate of decay from an excited $n$ to ground will be overestimated if the $l$-mixed transition probabilities are used. Another complication is that the relationships between the Einstein $A$ and $B$ coefficients are not correct if the total statistical weight of the upper configuration is used. All of these problems are solved with the $l$-resolved model now used in Cloudy. The current version of Cloudy predicts an H{\sc~i} spectrum in excellent agreement with \citet{SH95} when the number of resolved levels is made large enough.  

Predictions of the fluorescent excitation spectrum can have even greater differences. Table~\ref{tab:appendix} gives a comparison of the fluorescent-excited spectrum for an optically thin cell of pure hydrogen gas with unit density. It is exposed to a 50,000$K$ blackbody with an ionization parameter of 0.01. The emission would be close to Case A if there were no external continuum fluorescing the Lyman lines. It is in the Case C limit when continuum excitation is included. The first column gives the wavelength of a few Lyman, Balmer, and Paschen lines. The second column gives the \citet{SH95} Case A intensities relative to their predicted H$\beta$. The last two columns give the Case C predictions for the $l$-resolved (current Cloudy) and $l$-mixed (versions before C08) cases.

\begin{table}
\begin{center}
\caption{Comparison of predictions.\label{tab:appendix}}
\begin{tabular}{l|r@{.}lr@{.}lr@{.}l}
\tableline\tableline
 Line & \multicolumn{2}{c}{Case A} & \multicolumn{2}{c}{Case C ($l$-mixed)} & \multicolumn{2}{c}{Case C ($l$-resolved)} \\
\tableline
1216 &  33&1   & 146&    & 165&    \\
6563 &   2&82  &   4&02  &   8&02  \\
4861 &   1&    &   1&45  &   2&35  \\
4340 &   0&473 &   0&687 &   1&00  \\
4102 &   0&264 &   0&384 &   0&523 \\
1.875m & 0&455 &   0&496 &   0&807 \\
1.282m & 0&215 &   0&235 &   0&347 \\
\tableline
\end{tabular}
\end{center}
\end{table}

\end{document}